\documentclass[usenatbib,usegraphicx,useAMS]{mn2e}

\begin{document}

\title[Pattern speed estimates for 38 barred galaxies]{Model-based
  pattern speed estimates for 38 barred galaxies} \author[Rautiainen
  et al.]  {P. ~Rautiainen \thanks{E-mail:pertti.rautiainen@oulu.fi}
  H. ~Salo and E. ~Laurikainen\\ Astronomy Division, Department of
  Physical Sciences, University of Oulu, P.O. Box 3000,\\ FIN-90014
  Oulun yliopisto, Finland}

\date{Received x.x, accepted x.x}

\maketitle

\begin{abstract}
We have modelled 38 barred galaxies by using near-IR
and optical data from the Ohio State University Bright Spiral Galaxy
Survey. We constructed the gravitational potentials of the galaxies from
$H$-band photometry, assuming constant mass-to-light ratio. The halo
component we chose corresponds to the so called universal rotation
curve. In each case, we used the response of gaseous and stellar
particle disc to rigidly rotating potential to determine the pattern speed.   

We find that the pattern speed of the bar depends roughly on the
morphological type. The average value of corotation resonance radius
to bar radius, $\mathcal{R}$, increases from $1.15 \pm 0.25$ in types
SB0/a -- SBab to $1.44 \pm 0.29$ in SBb and $1.82\pm 0.63$ in SBbc --
SBc. Within the error estimates for the pattern speed and bar radius,
all galaxies of type SBab or earlier have a fast bar ($\mathcal{R} \le
1.4$), whereas the bars in later type galaxies include both fast and
slow rotators. Of 16 later type galaxies with a nominal value of
$\mathcal{R} > 1.4$, there are five cases, where the fast rotating bar
is ruled out by the adopted error estimates.

We also study the correlation between the parameter $\mathcal{R}$ and
other galactic properties. The clearest correlation is with the bar
size: the slowest bars are also the shortest bars when compared to the
galaxy size. A weaker correlation is seen with bar strength in a
sense that slow bars tend to be weaker. These correlations leave room
for a possibility that the determined pattern speed in many galaxies
corresponds actually that of the spiral, which rotates more slowly
than the bar. No clear correlation is seen with either the galaxy
luminosity or colour.    

\end{abstract}


\section{Introduction}
\label{introduction}

Studies in near-IR, where the extinction is lower than in visual
wavelengths and majority of light comes from the old stellar
population, have shown that about 60-70\% of all spiral galaxies have
a large scale stellar bar
\citep{eskridge2000,whyte2002,laurikainen2004,menendez2007,marinova2007}. According
to recent analysis of over 2000 spiral galaxies \citep{sheth2008}, the
bar fraction decreases from about 65\% in the local universe to about
20\% at redshift $z=0.84$ \citep[see
  also][]{abraham99,elmegreen2004,jogee2004,menendez2007}. Anyhow,
bars are so common that they are either very robust or they represent
a recurrent phenomenon in the life of a spiral galaxy
\citep{bournaud2005}. In contrast with some earlier studies
\citep{thompson81,elmegreen90c}, \citet{hernandeztoledo2007} claim the
bar frequency is roughly the same in different environments -- modest
interactions do not seem to play a major role in bar formation and
destruction.

Bars can be roughly divided into two classes based on their light
distribution: flat and exponential bars
\citep{elmegreen85,elmegreen96b}. In the first type, which is more
typical to early type barred galaxies, the radial surface brightness
profile along the bar major axis is flatter than in the surrounding
disc, whereas in the exponential bars the profile is quite similar to
the surrounding disc. Furthermore, the flat bars can display twisting
of isophotes \citep{elmegreen96a}. Bar morphology can also be either
classical or of ansae-type, characterized by blops at the both ends
of the bar \citep{laurikainen2007,martinez2007}.

Another approach to characterize a bar is its strength. Some attempts
have based on the ellipticity of the deprojected bar
\citep[e.g. ][]{martin95b,whyte2002,laurikainen2002b}. Recently, there
have been attempts to estimate the actual gravitational perturbation
of the bar by using near-IR photometry
\citep{buta2001b,laurikainen2002b,laurikainen2004,laurikainen2005,buta2005}. There
is also another bar strength estimate, namely the $A_2$-Fourier
amplitudes of density \citep{laurikainen2004,laurikainen2005}, which
is an approximation of the relative mass of the bar.  All these bar
strength estimates are discussed with respect to the Hubble sequence
by \citet{laurikainen2007}.

Perhaps the most important parameter defining a bar is its pattern speed,
$\Omega_{bar}$, or how fast the bar rotates. In principle, this
determines how far the orbits of stars and gas clouds are affected by
the bar. The pattern speed has a
physical upper limit -- a bar cannot reach beyond its corotation
resonance (CR) radius $R_{CR}$, i.e.\ the region in the disc where the
angular speed of circular rotation equals the bar
pattern speed. This limitation is based on the studies of stellar
orbits in barred potentials -- the orientation of the major axes of 
closed orbits becomes perpendicular with the bar beyond $R_{CR}$, thus
the orbits in the outer disc are not able to support the bar
\citep{contopoulos80a}. On the other hand, there is no evident lower
limit based on stellar orbits for the bar pattern speed.   

\begin{figure}
\resizebox{\hsize}{!}{\includegraphics{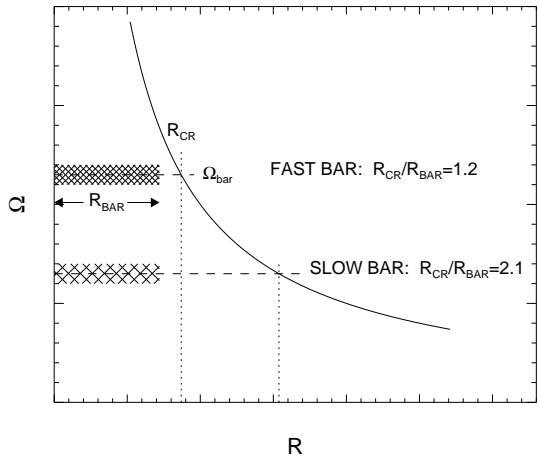}}
\caption{A schematic drawing of fast and slow bars.}
\label{barschema}
\end{figure}

In the literature an often used nomenclature
is based on the value of a dimensionless parameter $\mathcal{R} =
R_{CR} / R_{bar}$, where $R_{bar}$ is the semi major axis of the
bar. The cases where $\mathcal{R} \le 1.4$ are usually called ``fast bars'',
whereas those with a larger ratio are ``slow bars''
\citep{debattista2000}. This is shown as a schematic drawing in
Fig.~\ref{barschema}. Considerable effort has been devoted to
determine $\mathcal{R}$ for individual galaxies and to study if it
depends on other properties of the bar itself or the other galactic
components.     

One way to study the evolution of galaxies is to make self-consistent
N-body simulations, where the original particle distribution forms
different kinds of structures depending on its initial stage. If a bar
forms by a global bar instability in these simulations, it tends to
be fast rotating \citep{sellwood81}. On the other hand, it has been
suggested that if a bar forms in a galaxy due to an interaction with
other galaxies, it can be a slow rotator, perhaps extending only near
its inner Lindblad resonance \citep{miwa98}. This may also be the case
with gradual bar growth \citep{lynden79}, or when the initial
bulge-to-disc mass ratio is low \citep{combes93}. The
pattern speed of the bar does not depend only on its initial value --
the bar interacts with the other components of the galaxy. Secular
evolution due to interaction between the bar and the outer disc
decelerates the bar rotation, but this seems to be compensated by the
growth of the bar length by capturing more particles
\citep{sellwood81}. The situation is different with 
the interaction between the bar and the dark halo: if the halo density
is high in the inner parts of the galaxy, the resonant interaction can
decelerate an initially fast rotating bar so much that it becomes a
slow rotator \citep{debattista98,athanassoula2003}. Thus, the observed
bar pattern speed could be considered as a rough estimator for the
halo contribution. 

Much weight has been given to a model-independent kinematic method,
the so-called Tremaine--Weinberg method, hereafter TW
\citep{tremaine84}. It is based on photometric and spectroscopic
observations along slits parallel to the galaxy's major axis and assumes a
rigidly rotating component, which satisfies the continuity
equation. TW-method has been used to determine pattern speeds of about
20 large scale stellar bars. The results range from probably unphysical
$\mathcal{R}=0.6$ to slow bar regime with $\mathcal{R}=1.8$. This has
been considered to be in accordance with fast bars, at least when the
quite large error estimates \citep[e.g.\ due to sensitivity to errors
  in the position angle of the disc, ][]{debattista2003} are taken into account
\citep*{kent87,merrifield95,gerssen99,gerssen2003,aguerri2003,debattista2004,corsini2007,treuthardt2007}.
This apparent lack of slowly rotating bars has been interpreted
to indicate that the density of a dark halo is small in the inner
parts of barred galaxies \citep{debattista2000,aguerri2003}, which is
in disagreement with some cosmological simulations of structure
formation, typically producing centrally peaked dark matter component
\citep{navarro96}. However, due to limitations set by the initial
assumptions, measurements using stellar absorption line spectrometry
have been made almost exclusively for SB0-galaxies, which are almost
free of gas and dust. 

Although some attempts \citep{zimmer2004,rand2004} have been made to
use TW-method with CO observations, the pattern speeds of late type
barred galaxies have been usually determined by various indirect methods.
Many of these rely on morphological features in barred galaxies such
as spiral arms, rings or leading offset dust lanes. In a classic
scenario, a two-armed spiral starts from the ends of the bar, but
exceptions to this are not rare: there can be an ``empty'' region
between the bar
and the spiral (e.g.\ NGC 210) or an offset between the bar
position angle and the starting points of the spiral arms (e.g.\ NGC
799). There can also be more than two arms either in the
whole scale of spiral structure (e.g. ESO 566-24) or as
a multiply armed outer disc (e.g.\ NGC 4303). The clearest
correlation with the bar properties is observed in the sizes and shapes of
rings \citep[for a detailed review, see ][]{buta96b}: outer
rings whose semi major axis is about twice the bar radius and inner
rings whose major axis is usually the same as the major axis of the
bar. There is also a third ring type, the nuclear rings, whose radius
is roughly 1/10 of the bar radius, but with a large scatter. Leading
offset dust lanes are straight or curved dust features inside the
bar and are located in its leading side. 

Models of individual galaxies have been constructed either by fitting
analytical potential components to observations or by determining the
potential from photometry, making some assumption about
mass-to-luminosity ratio, orientation and internal geometry of galaxies. 
In principle, modelling is more economical than using TW-method - the
pattern speed can be estimated by comparing the simulated and observed
morphology. Naturally, kinematical observations can help to determine
the acceptable parameter range. Modelling has
produced both high and low pattern speed estimates for the bars
\citep{hunter88,sempere95a,lindblad96b,salo99,rautiainen2004}. The
given error estimates of models are typically smaller than with TW-method.

Pattern speed estimates have also been done by identifying various
morphological or photometric features with resonances. Inner and outer
rings are usually considered to form by gravitational torque of the rotating
bar, which causes gas to flow in the radial direction. The net torque
vanishes in major resonances, where gas then accumulates
\citep{buta96b}. Comparison with results of gas dynamical simulations
and analysis of orbits in barred galaxies have led to the following
identifications: an outer ring should be located near the Outer Lindblad
resonance (OLR) and the inner ring near the inner 4/1-resonance, which
is located inside $R_{CR}$
\citep{schwarz81,schwarz84b,byrd94,salo99,rautiainen2000,patsis2003}. When
the rotation curve is known, these resonance identifications can be used
to determine the pattern speed \citep{buta98a}. Another approach is
based on two-colour photometry \citep{puerari97}: the location of
shock induced star forming regions should change side related to
spiral arms when crossing $R_{CR}$. This crossing should be detectable
from two-colour photometry. Especially, the results of
\citet{aguerri98} gave a hint (the total sample consisted of 10
galaxies) that late type spirals could be slower rotators than early types.  

In addition there are at least two other
methods worth mentioning. A kinematical method to determine the
corotation radius, which is based on the residual patterns in the
velocity field after removal of circular velocities, was suggested by
\citet{canzian93}. However, it has been rarely applied. Recently,
\citet{zhang2007} suggested that the calculated phase shifts between
potential and density could be used to determine $R_{CR}$. The
validity of this approach is still questionable: for several galaxies
it found corotation radii well inside the bar, which is in
disagreement with the analysis of orbits in barred potentials. 

Using the previously published pattern speed estimates in studying a
possible dependency on the morphological type is problematic. Very
different methods have been used, the data is of uneven quality or the
definition of the bar differs. There are several papers where quite
successful simulation models are presented, but only one pattern speed
has been tried (e.g.\ by assuming $\mathcal{R}=1$) -- this cannot be
taken as a {\it determination} of the pattern speed.

Large galaxy surveys and the increase of computing power makes it
possible to improve the situation by mass production of galaxy models:
here we present a simulation series where we estimate the pattern
speeds of 38 moderately inclined barred galaxies, using data 
from the Ohio State University Bright Spiral Galaxy Survey
\citep[][hereafter OSUBSGS]{eskridge2002}. To our knowledge, this is the
largest sample of barred spiral galaxies whose pattern speeds are
determined with a consistent method. Furthermore, the
morphological types of these galaxies range
from SB0/a to SBc, based on Third Reference Catalogue of Bright
Galaxies \citep[][hereafter RC3]{devaucouleurs91}, so that each
morphological type is represented by several galaxies. The initial
modelling results for this sample were published in \citet[][ hereafter
  RSL2005]{rautiainen2005}, and part of the analysis in this paper was
also presented in \citet{salo2007}. 

Whereas we find all the bars of early type galaxies 
(SBab or earlier) of the sample are fast rotators, our models
also suggest that galaxies of later morphological types include both fast
and slow rotating bars. The slow rotation seems to be related to the
small size of the bar. An alternative interpretation to this is that
the spiral arms in these galaxies have a lower pattern speed than the
bar, a situation often seen in N-body models \citep{sellwood88,rautiainen99}.

\section{Modelling}
\label{Modelling}

Our modelling method is essentially the same as we used in modelling
IC 4214 and ESO 566-24 \citep{salo99,rautiainen2004}, the main
difference being that we do not have kinematical data for this larger
sample. We assume that the $H$-band light distribution corresponds to
the projected distribution of luminous matter (constant $M/L$) and
that there is only one pattern speed in each galaxy.

We simulated the behavior of two-dimensional discs of collisionless
stellar test particles and inelastically colliding gas particles in
the determined potentials. In each collision, the relative velocity
component of two gas particles in the direction joining the particle
centres was reversed and multiplied by the coefficient of restitution,
here assumed to be zero. The main modelling parameter was the pattern
speed of the non-axisymmetric part of the potential. For more details
on the simulation code, see \citet{salo91a} and \citet{salo99}.  The
best fitting pattern speed and a rough estimate of the error (the
average error estimate for $R_{CR}$ is about $\pm 15$\%) was
determined by visual comparison between the simulation morphology and
the morphology in the $B$- and $H$-band images.

\subsection{Mass model}

We used the $H$-band images from the OSUBSGS,
for which a bulge--disc--bar decomposition was made by
\citet{laurikainen2004}. This multiple-component process gives more
realistic bulge parameters than fitting just disc and bulge. For each
image the bulge component was first removed, after which the disc was
deprojected to face-on orientation. The light distribution and surface
density of the disc was approximated by a Fourier decomposition
using the same approach as in \citet{laurikainen2002b}. 

The disc gravity was
calculated using even azimuthal components from m=0 to 8. The
strongest of the odd components is usually m=1. However, it can
represent lopsidedness caused by a recent interaction and even if it
is related to an internal mode, it can have a different pattern speed than the
m=2 mode. The omission of odd modes can cause some of the
models look more regular than the observed galaxy.

\begin{figure}
\includegraphics[width=\linewidth]{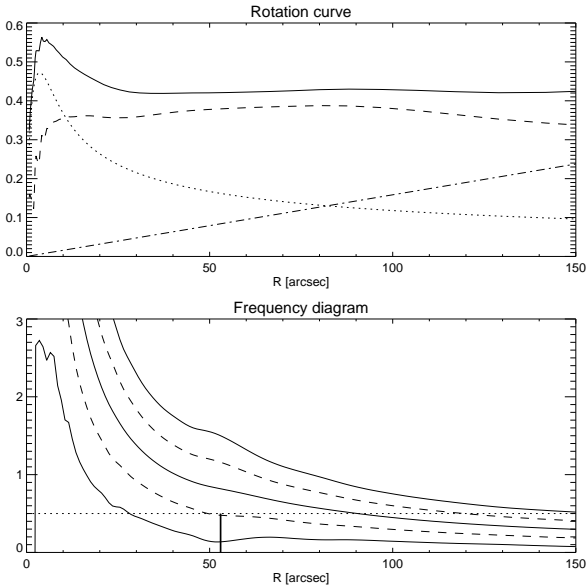}
\caption{The calculated rotation curve (top) and corresponding
  frequency diagram (bottom) for NGC 4303. The continuous line in the
  top frame shows the rotation curve, the dotted line the bulge
  contribution, the dashed line the disc and dot-dashed line the halo
  contribution. In the bottom frame the curves drawn with continuous
  line corresponds to $\Omega$ and $\Omega \pm \kappa/2$, the dashed
  lines corresponds to $\Omega \pm \kappa/4$ and the the dotted line
  indicates the best fitting pattern speed. Also the bar radius
  $R_{bar} \approx 53 \arcsec$ is indicated. This close coincides with
  inner 4/1-resonance.}
\label{rota_example}
\end{figure}

In the vertical direction, an
exponential distribution was assumed, with a constant scale
height throughout the disc: the scale height was chosen according to
the morphological type of the galaxy, using the typical ratio of
vertical to radial scale lengths $h_R$ given by \citet{degrijs98}.  The
gravitational potential of the bulge was added to the disc potential,
assuming that the bulge mass is spherically distributed. In our standard
mass model we also included a dark halo component based
on the universal rotation curve of \citet{persic96}, in a
similar manner as in \citet{buta2004}. To calculate the halo profile,
we used $L/L_\star$ calculated from the RC3 and NED (NASA/IPAC
Extragalactic Database) data and distance from \citet{tully88} (except
for NGC 6782 for which the distance was calculated from the radial
velocity assuming Hubble constant $H_0=75 \ \mathrm{km \ s}^{-1}
\ \mathrm{Mpc}^{-1}$). For more details of the potential calculation
see \citet{laurikainen2002b}. An example of a calculated rotation
curve and corresponding frequency diagram showing $\Omega$ (circular
frequency) and $\Omega \pm \kappa/2$ (where $\kappa$ is the epicycle
frequency), is shown in Fig.~\ref{rota_example}.  

Because we do not have kinematical data, we cannot fine-tune our mass
models. However, we can estimate the robustness of our pattern speed
fits by trying different mass models.  In addition to our standard
model with universal rotation curve, we also made simulations without
the dark halo component. In general, the calculated rotation curves
were then falling in the outer parts of the discs, but the determined
pattern speeds were quite similar because the fitting was done mostly
using features in the inner parts. For eight galaxies (six with slow
bars, two with fast bars), we also studied the effect of increasing
the halo contribution. This was tested by multiplying the even Fourier
amplitudes $m=2 - 8$ of the force by a factor 0.75 or 0.5. The effect
to the fitted pattern speed was quite small, in all cases within the
error estimates of the standard model series. In a couple of cases,
the fit was marginally better when the amplitudes were multiplied by
0.75, but in all cases the fit was worse with multiplication factor
0.5.  This is in good agreement with our tests with the mass models
for ESO 566-24 \citep[see Figures 4 and 12 in ][]{rautiainen2004},
where we found that three mass models (no halo, a halo closely
corresponding to universal rotation curve and dominating halo) all
gave essentially the same pattern speed, although the quality of the
fit was clearly different. Also, modifying the mass model so that the
major axis rotation curve determined from the simulated velocity field
matched the observed one as closely as possible, changed the
corotation resonance distance only slightly.

\subsection{Bar radius}

There is no unambiguous way to determine the bar radius \citep[for a
  detailed analysis see e.g. ][]{athanassoula2002a}. Some approaches
rely on Fourier analysis of the surface brightness: the bar is
identified with the region where the $m=2$ (or $m=4$) phase angle is almost
constant, or where the amplitude drops to a certain fraction of its
maximum value. Another method is to fit
ellipses to isophotes -- bar radius is then determined as the radius where
the ellipticity reaches a maximum or where there is a steep drop in
it. Besides these, there are also several other methods.

In RSL2005 we estimated lengths of the bars by using visual inspection
of deprojected H-band images and isophotes and $m=2$ Fourier phase
angles. In this study,we have basically followed the same approach as
\citet{erwin2004,erwin2005} \citep[see also][]{michel2006}, with
slight modifications. We first remove the surface brightness of the
fitted bulges from the images and then deproject them. Next ellipses
are fitted on the isophotes of the deprojected images by using routine
ELLIPSE in IRAF. For the lower limit of the bar radius we normally use
the semimajor axis of maximum ellipticity $a_\epsilon$ (the outermost
maximum that is still within the bar). As the upper limit we use
$L_{bar}$ which is the minimum of two ellipse-fit measures -- the
first minimum ellipticity outside $a_\epsilon$, or the point where the
position angles of the fitted ellipses differ more than 10$^\circ$
from the position angle of the bar (the smaller one of these is
chosen). If a galaxy have an inner ring that interferes the
determination of $a_\epsilon$ and/or $L_{bar}$, we use the inner edge
of the ring as the lower limit and its outer edge as the upper limit
for the bar radius, both measured along the bar major axis. The bar
radius is then determined as $R_{bar}=(a_\epsilon + L_{bar})/2$ and
its error estimate is taken as $\Delta{R_{bar}} =
R_{bar}-a_\epsilon=L_{bar}-R_{bar}$.  The adopted values are given in
Table 1, which also lists the error estimates. The error
estimates in the bar radius are in range $\pm$4\% -- $\pm$24 \%, the
average being about $\pm$14\%.

When comparing our results with other bar length measurements, we
limit to those done with near-IR photometry.  The bar lengths used in
this study are on average about 10\% longer than the values given by
\citet{laurikainen2004}, who used the phase angle method. This
typically corresponds to our lower limits for the bar lengths. In five
cases our bar radius estimate is clearly smaller compared to
\citet{laurikainen2005}: the difference is related to the
identification of the bar: for example if a galaxy has an outer ring
of type $R_1$, then the innermost part of the ring is close to the
major axis of the bar, and can make the $m=2$ phase angle almost
constant far beyond the actual bar region. A good example of this is
NGC 3504.

Eleven of our galaxies are common with \citet{erwin2004,erwin2005},
who used essentially the same method to determine the bar
radius. Their bar radius estimates (taking into account possible
differences in the adopted orientation parameters) are in good
accordance with our results: typically our bar radius estimate is
between their values for $a_\epsilon$ and $L_{bar}$, and if that is
not the case, the difference is only marginal. The only exception of
this is NGC 4303, for which we get 60 \% higher value.

\citet{marinova2007} used the OSUBSGS images in their study of bar
properties. They give bar radii based on isophote ellipse fitting for
34 galaxies that are also in our sample. Typically our bar radius
estimates are few percent larger than their values (median is about
5\%), which is not surprising taking into account that this is
expected to correspond to our lower limit. The difference is largest
for NGC 1317 where \citet{marinova2007} gives measurements for a
feature we consider to be a secondary bar or a nuclear ring. It is the
only case in their sample, where the given bar radius is not within
our error estimates.  In addition to previous examples, three of the
galaxies are common with \citet{laine2002} (who used the maximum of
fitted ellipticity) and one with \citet{reese2007} (distance where
10\% of the maximum of fitted bar intensity profile is reached). Our
bar radius estimates agree well with their results when error
estimates are taken into account.

\subsection{Defining the best fit}

The morphological features we used in the comparison are typically the
spiral arms, the inner rings which surround the bar, and the outer
rings. With spiral arms we compared their extent, location and pitch
angles, and occasionally other features such as straight arm
segments. With rings we compared the shapes and sizes to the observed
ones. This comparison was done with the help of a crosshair figure
(cross + circles indicating the disc scale length and its multiples),
that was overplotted on both the galaxy images and the modelled gas
and stellar particle distributions.

\begin{figure*}
\resizebox{\hsize}{!}{\includegraphics{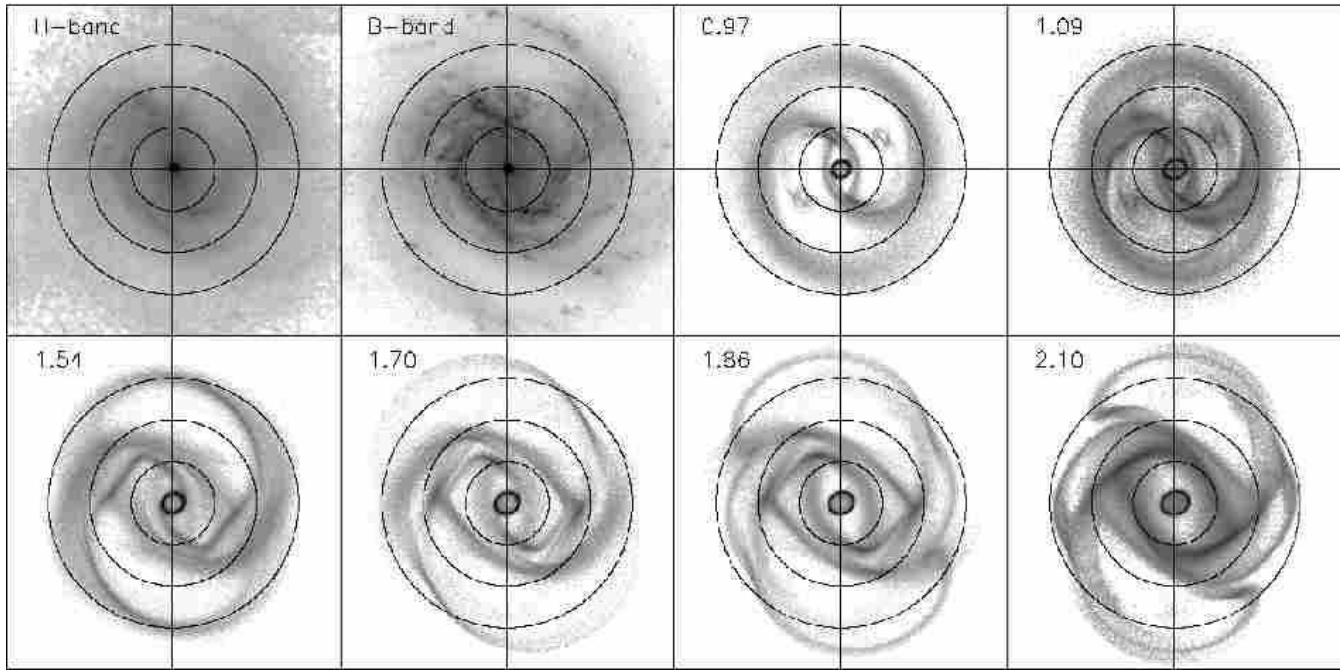}}
\caption{Models with different pattern speed for NGC 4303. First two
  frames shows deprojected images of the galaxy in $H$- and
  $B$-bands. The number in the simulation frames indicates
  $\mathcal{R} = R_{CR}/R_{bar}$. The radii of the drawn circles
  corresponds to 1, 2 and 3 scale lengths of the disc in the
  $H$-band.}
\label{ngc4303_pspeeds}
\end{figure*}

An example of our modelling approach is shown in
Fig.~\ref{ngc4303_pspeeds}, where we compare models with six different
pattern speeds to $H$- and $B$-band images of NGC 4303. In a few cases
we could not determine a single best-fitting pattern speed because the
outer parts favored lower pattern speed than the inner parts. A
possible reason for this is that a galaxy might have many pattern
speeds so that the spiral rotates more slowly than the bar. This
is often seen in N-body simulations of barred galaxies
\citep{sellwood88,masset97,rautiainen99}. If two fits are considered to
be of equal quality, the value adopted in this study
corresponds to the smaller $\mathcal{R}$ (faster bar), best characterizing
the bar region. The ambiguity of pattern speed determination was also
faced by \citet{lindblad96b}: for galaxy NGC 1300 they found two
values giving $\mathcal{R} \approx 1.3$ or $\mathcal{R} \approx 2.4$,
both producing equally good fit.

Several of the galaxies contain leading offset dust lanes in their
bars. Although such features are sometimes seen also in our models
(see Fig.~\ref{ngc4303_pspeeds}), we did
not include them to pattern speed determination: the dust lanes
correspond to highest density in the interstellar medium and it is
probable that our gas modelling method \citep{salo91a}
is not suitable for such situations. 

Most of our models contain features in their gas components that look
like nuclear rings observed in galaxies. Although several studies have
connected nuclear rings to inner Lindblad resonance (ILR) or
resonances, it has been claimed, based by gas dynamical models, that
they are actually not related to resonances but to orbits in the
central part of the bar \citep{regan2003}. Furthermore, in the models
of Regan \& Teuben the nuclear rings shrank even when the
gravitational potential was constant. Even so, in our previous models
of IC 4214 and ESO 566-24, the model which gave the best fit to
overall morphology and kinematics, produced also a nuclear ring whose
major axis diameter was the same with that of the observed ring. However,
we have not included the nuclear rings to comparison with the sample
galaxies: the image quality is usually not good enough to show a
possible nuclear ring and the gravitational potential, calculated from
the $H$-band image, is probably not accurate enough in the innermost
parts due to low resolution (one pixel corresponds to 0.05 -- 0.2 kpc)
Also, some of the galaxies included active galactic nuclei or central
star bursts, which affect the central parts of the brightness profiles
even in the $H$-band, and thus also the deduced potentials \citep[see
  Figs. 2 and 3 in][]{salo99}.   

In addition to gaseous nuclear rings, our models often show distinct
stellar features, whose major axis is perpendicular to the bar, in the
same region. These features, which are usually not seen in real
galaxies, are related to $x_2$-orbits. \citet{patsis2005} demonstrated
that $x_2$-features can be removed from the collisionless test
particle models by fine-tuning the initial conditions. On the other
hand, this region often has a secondary bar of about similar size as a
typical nuclear ring but a different pattern speed than the main bar
\citep[e.g.][]{friedli93b,rautiainen99,maciejewski2000,heller2001,rautiainen2002,maciejewski2006}.

\begin{figure*}
\resizebox{\hsize}{!}{\includegraphics{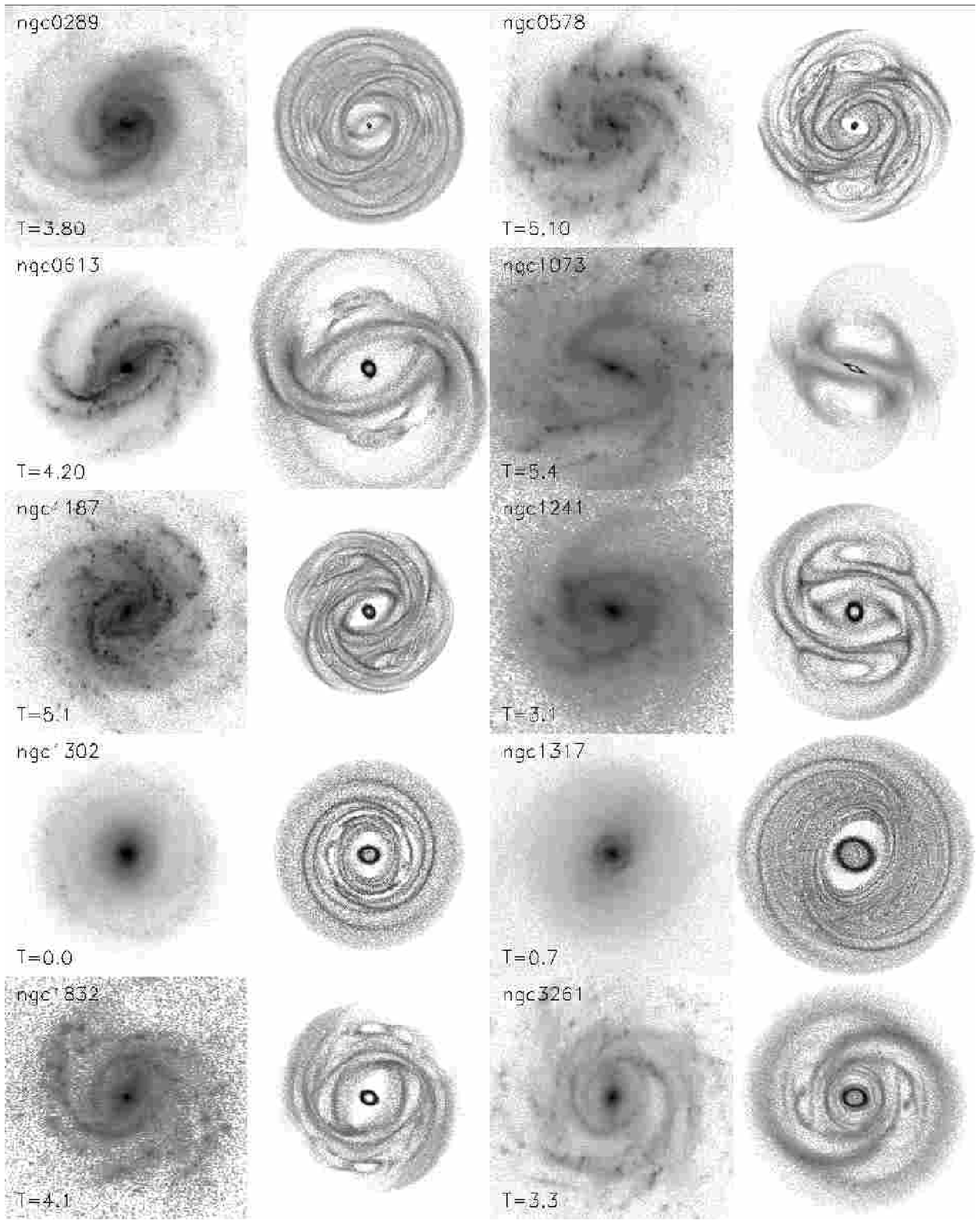}}
\caption{Deprojected image of the sample galaxies ($B$-band)
  vs. models (gas or stellar component). The Hubble stage $T$ is also
  given.}
\label{modsel1}
\end{figure*}

\begin{figure*}
\resizebox{\hsize}{!}{\includegraphics{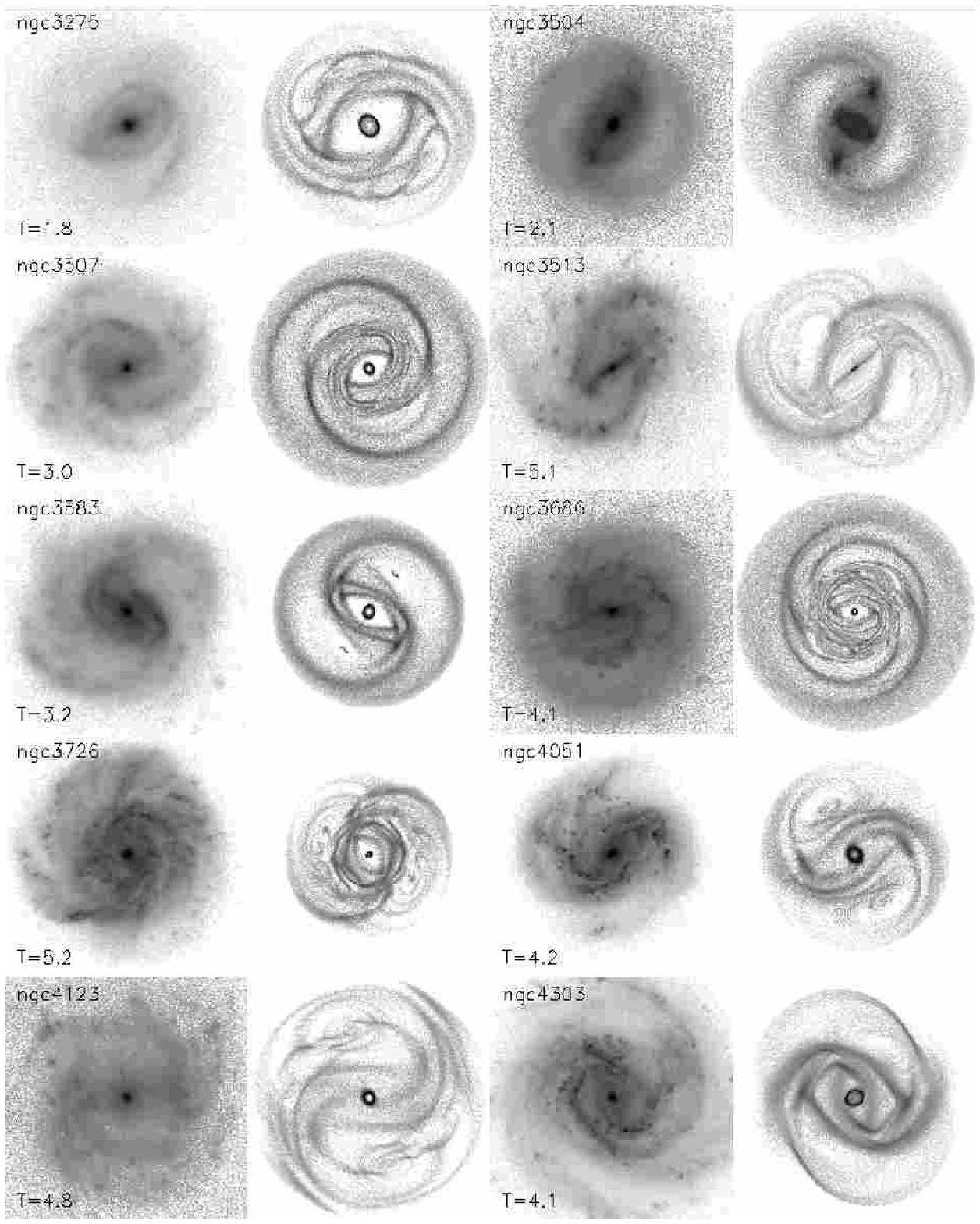}}
\contcaption{}
\end{figure*}

\begin{figure*}
\resizebox{\hsize}{!}{\includegraphics{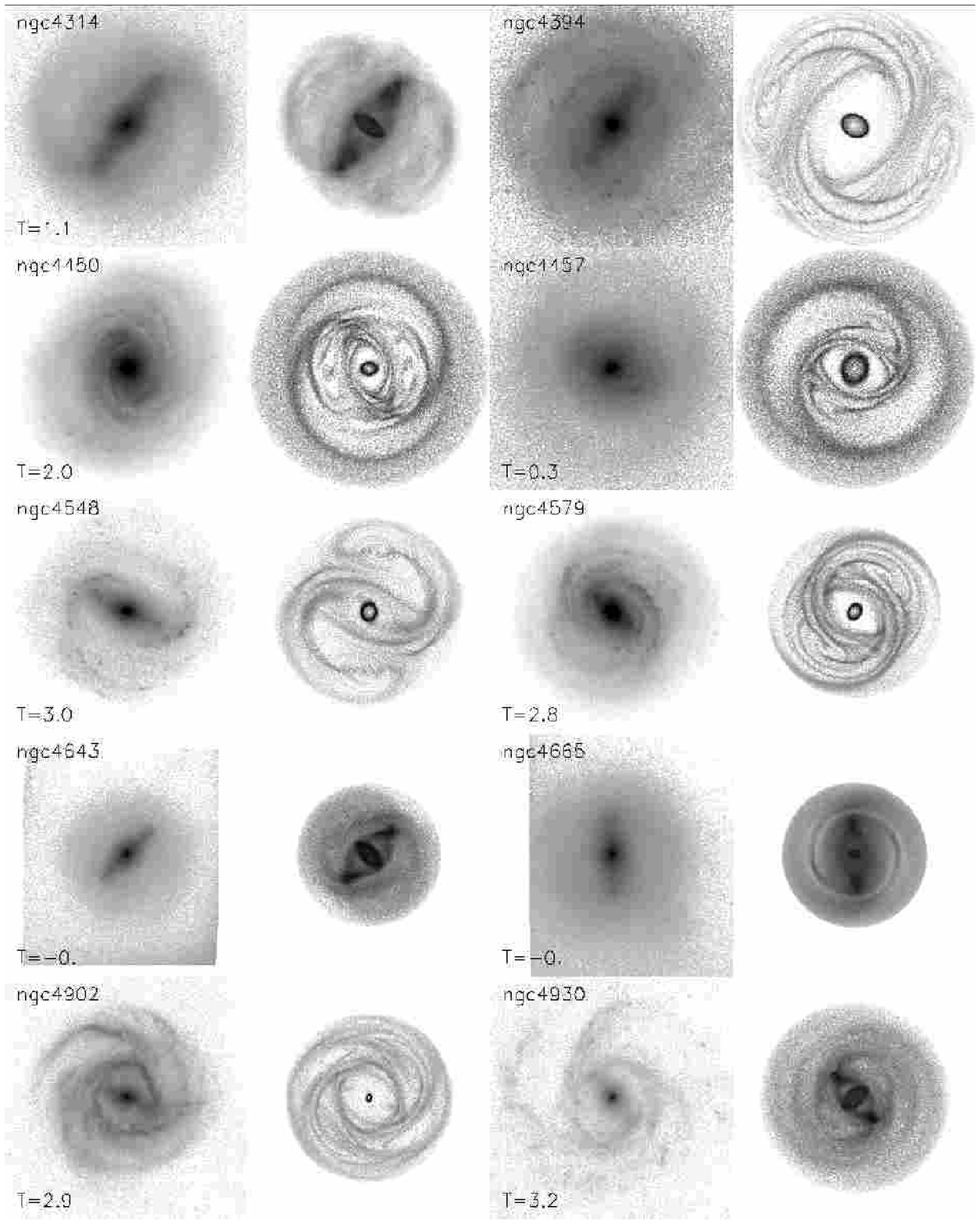}}
\contcaption{}
\end{figure*}

\begin{figure*}
\resizebox{\hsize}{!}{\includegraphics{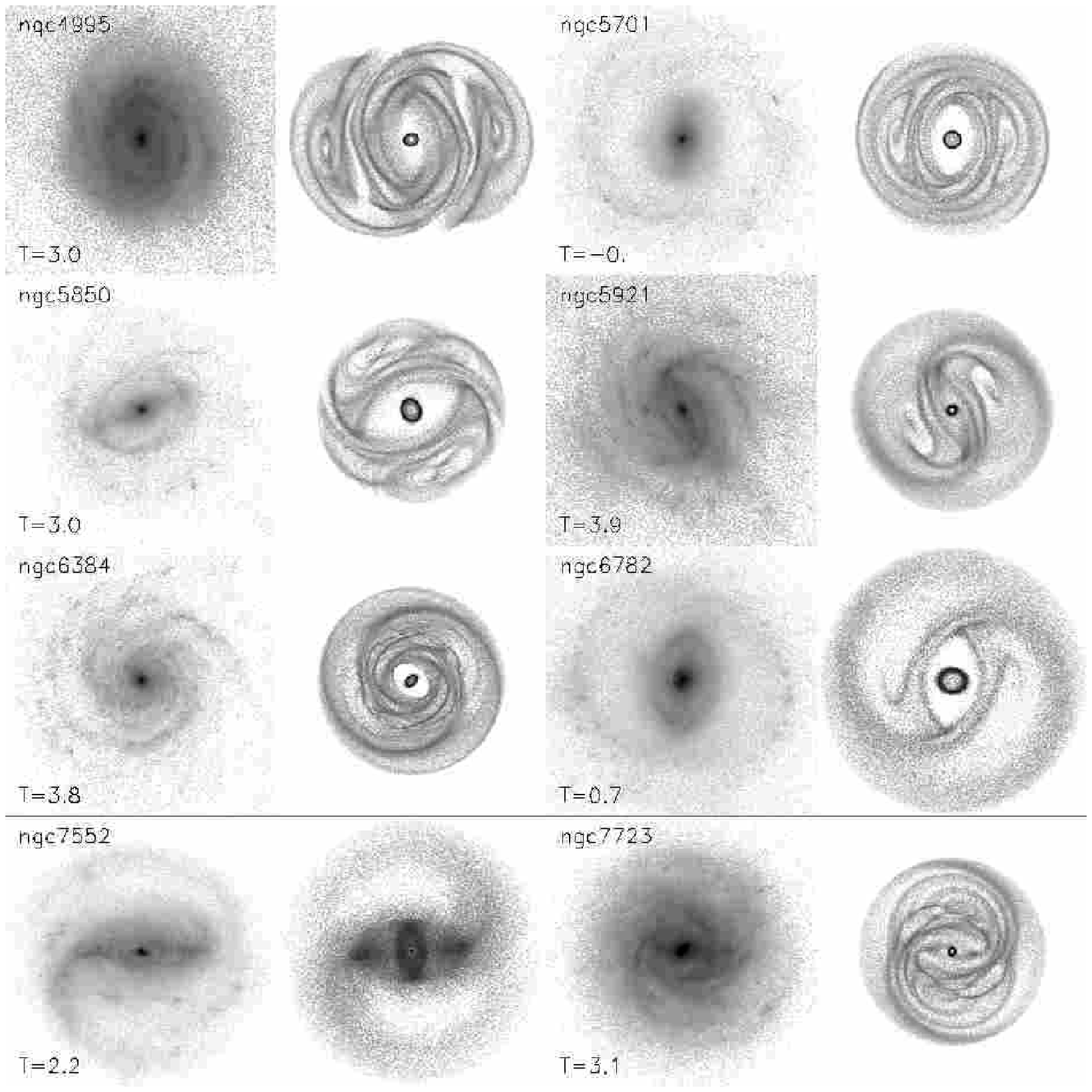}}
\contcaption{}
\end{figure*}

\section{Notes on individual galaxies}

Here we give comments on some of the modelled galaxies.

{\bf NGC 289}: this SBbc ($T=4$) galaxy has a small bar, the inner
spiral arms start from the ends of the bar. There is also an extended
outer spiral structure, which bifurcates from the inner spiral
arms. The best-fitting model, which sets $\mathcal{R} \approx 2.6$
reproduces the inner spiral, the bifurcations and the inner part of
the outer spiral (see Fig.~\ref{modsel1}). The outermost part of the
outer spiral structure is not reproduced, but it can be related to an
interaction with a small companion galaxy, LSBG F411-024. 
model sets $R_{CR}$ much outside

{\bf NGC 613}: this SBbc ($T=4$) galaxy has a bar, inner ring and
several spiral arms emerging from the ring. The Fourier phase angle
method gives too high bar radius, well in the area of spiral
arms. The fit, which sets $\mathcal{R} \approx 1.6$, is not very
good (see Fig.~\ref{modsel1}): only the inner ring and small parts of
the arms are reproduced. The inner ring would allow also a bit higher pattern
speed, which would set the galaxy clearly to the domain of fast bars.

{\bf NGC 1187}: this SBc ($T=5$) galaxy has a bar with three spiral
arms emerging from it. The best-fitting model (see
Fig.~\ref{modsel1}), setting $\mathcal{R} \approx 2$, has four arms
due to using even Fourier components,
three of which corresponds well the observed arms. The error estimate
for $\mathcal{R}$ is quite large, and a fast-bar solution cannot be ruled out. 

{\bf NGC 1241}: in this SBb ($T=3$) galaxy the bar is surrounded
by an inner pseudoring, which exhibit rectangular features, the spiral
structure is not regular, possibly due to interaction with
NGC 1242. The best-fitting model (see Fig.~\ref{modsel1}), setting $\mathcal{R}
\approx 1.4$, reproduces well the rectangular structure of the inner
pseudoring and also part of the spiral structure. 

{\bf NGC 1302}: the only distinct feature in this SB0/a ($T=0$)
galaxy is the very tightly wound spiral structure. The best-fitting
model (see Fig.~\ref{modsel1}), setting $\mathcal{R} \approx 1.6$, is
chosen by the extent of the gas spiral. 

{\bf NGC 1832}: this SBbc ($T=4$) galaxy has an inner ring whose
major axis seems to be larger than the bar. The spiral structure
asymmetric. The ring and one spiral arm are reproduced by the
best-fitting model (see Fig.~\ref{modsel1}), setting $\mathcal{R}
\approx 1.7$. 

{\bf NGC 3275}: this is a SBab ($T=2$) galaxy with an inner ring
and asymmetric spiral structure. The ring and part of the spiral
structure is reproduced with $\mathcal{R} \approx 1.5$ (see
Fig.~\ref{modsel1}). The Fourier phase angle method clearly overestimates the
bar radius: it is outside the lopsided inner ring. 

{\bf NGC 3504}: this is a SBab ($T=2$) galaxy with an outer
ring. The ring and the extent of the bar is best reproduced with a
simulation of stellar test particles giving $\mathcal{R} \approx
1.2$ (see Fig.~\ref{modsel1}). In the gas component the same pattern
speed is favored, but the fit is worse because the outer ring does not
form. The outer pseudoring in this galaxy causes the Fourier phase angle
criterion to work incorrectly, giving too high value for bar length. 

{\bf NGC 3507}: this SBb ($T=3$) galaxy has really long spiral
arms that can be followed about 360$^\circ$. The best-fitting model,
setting $\mathcal{R} \approx 1.2$, reproduces the spiral structure in
its full extent. Overall, this is one of the best models (see
Fig.~\ref{modsel1}). 

{\bf NGC 3583}: this SBb ($T=3$)galaxy has an inner pseudoring
with rectangular features and two-armed spiral. The best-fitting
model, with $\mathcal{R} \approx 1.2$, reproduces the shape of the
pseudoring and also the spiral structure is well fitted (see
Fig.~\ref{modsel1}). 

{\bf NGC 3726}: this SBc ($T=5$) galaxy has a weak and small
bar. The inner part of the spiral structure is somewhat
polygonal. The best-fitting model reproduces this appearance and
gives $\mathcal{R} \approx 2$ (see Fig.~\ref{modsel1}).

{\bf NGC 4303}: this SBbc ($T=4$) galaxy has a small bar inside
far extending spiral structure whose inner part shows straight
arm segments, sometimes called ``Vorontsov-Velyaminov rows''
\citep{chernin2000}. This structure is reproduced with a narrow
pattern speed range (see Fig.~\ref{ngc4303_pspeeds}) by a model giving
$\mathcal{R} \approx 1.7$.

{\bf NGC 4548}: this SBb ($T=3$) galaxy has spiral arms that seem
to start offset to the bar in the leading side -- at least the part
connecting spiral to the ends of the bar is very weak. The best-fitting model
gives $\mathcal{R} \approx 1.3$ (see Fig.~\ref{modsel1}). One arm of
this galaxy shows straight segments, unlike the case of
NGC 4303, these are not reproduced by the model.

{\bf NGC 5701}: the outer pseudoring of this SB0/a ($T=0$) galaxy
is almost detached from the bar. It is reproduced with a model giving
$\mathcal{R} \approx 1.4$ (see Fig.~\ref{modsel1}). 

{\bf NGC 5850}: the inner ring of this SBb ($T=3$) galaxy is
reproduced with a model giving $\mathcal{R} \approx 1.4$ or $R_{CR}
\approx 105 \pm 12 \arcsec$ (see Fig.~\ref{modsel1}). The outer
structure of this galaxy is not regular, possibly due to interaction
with NGC 5846 \citep{higdon98}. A hydrodynamical model
for this galaxy has been published by \citet{aguerri2001}, giving
$R_{CR} \approx 90 \arcsec$. We found that an
alternative model, giving $R_{CR} \approx 82 \pm 6 \arcsec$ or
$\mathcal{R} \approx 1.1$ is also possible when matching the size of
the inner ring, but then the shape is not correct. 

{\bf NGC 6384}: this SBbc ($T=4$) galaxy has a small bar and long
extending spiral structure that winds about 360$^\circ$. The
best-fitting model, giving $\mathcal{R} \approx 2.3$ reproduces the
spiral in its full length (see Fig.~\ref{modsel1}).

{\bf NGC 6782}: the inner ring and the large part of the
spiral of this SBa ($T=1$) galaxy are reproduced by a model giving
$\mathcal{R} \approx 1.3$ (see Fig.~\ref{modsel1}). The main
difference between the observed and modeled morphology is that the
real galaxy has an $R_1'$ outer pseudoring whereas the spiral in the
model does not form the pseudoring. The Fourier phase method clearly
overestimates the bar radius by setting it  well outside
the inner ring. This is possibly caused by
the outer pseudoring that affects the phase angle of the Fourier
decomposition. 

\begin{figure*}
\resizebox{\hsize}{!}{\includegraphics{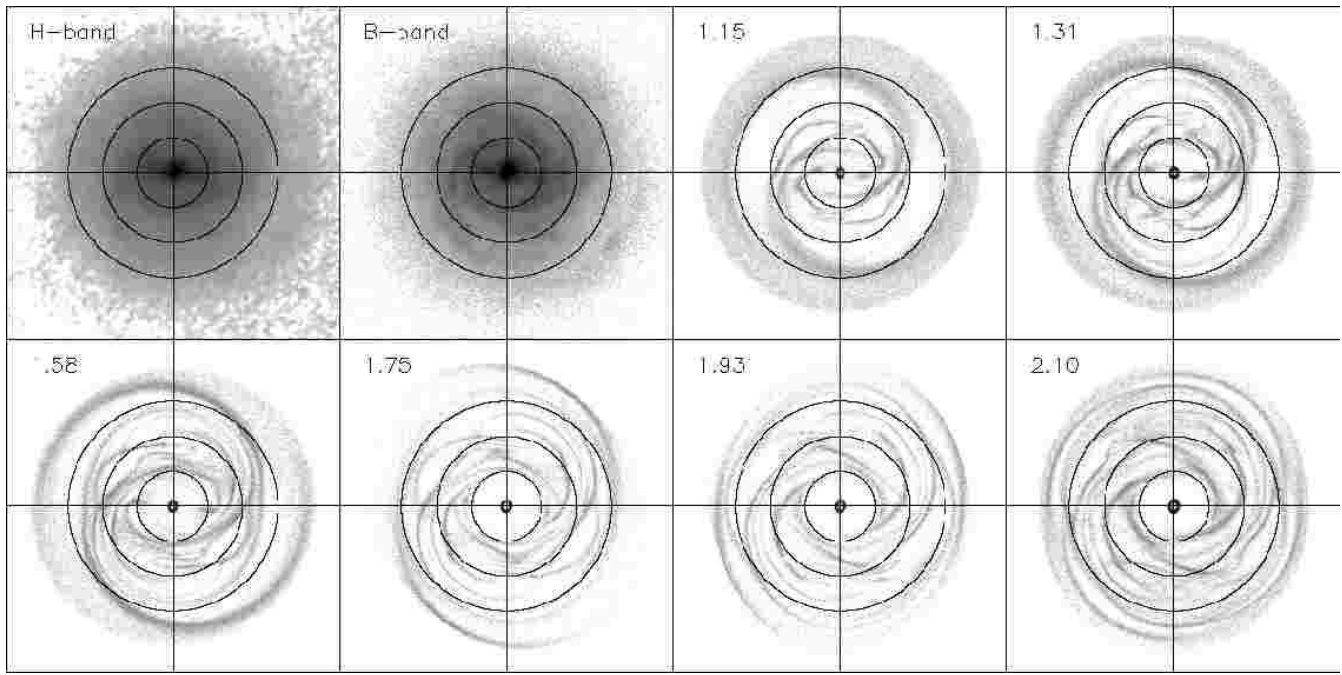}}
\caption{Models with different pattern speed for NGC 7723. First two
  frames shows deprojected images of the galaxy in $H$- and
  $B$-bands. The number in the simulation frames indicates
  $\mathcal{R}$. The radii of the drawn circles corresponds to 1, 2
  and 3 scale lengths of the disc in the $H$-band.} 
\label{coro7723}
\end{figure*}

{\bf NGC 7723}: the inner pseudoring and multiple-armed
spiral of this SBb ($T=3$) galaxy are best reproduced by a model giving
$\mathcal{R} \approx 1.3$. There is an alternative model, with $\mathcal{R}
\approx 1.9$, which gives about as good fit with the spiral
structure. For comparison see Fig.~\ref{coro7723}. 

\section{Results and discussion}

The results -- the bar radii and the fitted corotation resonance
distances are listed in Table 1, along with error
estimates for both parameters. The differences to preliminary values
for $R_{CR}$ that were given in RSL2005 are minor, within the error
estimates. With some galaxies the values of $\mathcal{R}$ has changed
considerably due to different way to determine the bar
radius. However, in the sample level this does not introduce any
significant changes in the observed dependency e.g. on the
morphological type.

In addition to $\mathcal{R}$ vs.\ Hubble stage comparison that was
already done for the preliminary values in RSL2005, here we search
possible correlations between the determined pattern speed and the
size and the strength of the bar. We also study if the fitted
$\mathcal{R}$ depends on the bulge contribution, galaxy luminosity or
colour.

\setcounter{table}{0}
\begin{table*}
\begin{minipage}{126mm}
\label{ometable}
\caption{The measured bar length $R_{bar}$ and its error estimate $\Delta
  R_{bar}$, the defined corotation radius $R_{CR}$ and its error
  estimate $\Delta R_{CR}$for each galaxy. Also given are parameter
  $\mathcal{R}$, error estimate $\Delta \mathcal{R}$ and Hubble stage
  $T$ (from HyperLeda).}
\begin{tabular}{lcccccccc}
\hline Galaxy & $R_{bar}$ & $\Delta R_{bar}$ & $R_{CR}$ &
$\Delta R_{CR}$ & $\mathcal{R}$ & $\Delta \mathcal{R}$ & $T$\\
& (arcsec) & (arcsec) &
(arcsec) & (arcsec) &  & \\  
\hline

NGC 0289 &   23.8 &    2.2 &   61.0 &    9.5 &   2.56 &   0.58 &  3.8\\
NGC 0578 &   23.0 &    2.5 &   78.9 &   15.8 &   3.43 &   0.95 &  5.1\\
NGC 0613 &   78.6 &    3.2 &  126.2 &   14.6 &   1.61 &   0.24 &  4.2\\
NGC 1073 &   51.0 &   10.5 &   48.7 &    4.4 &   0.95 &   0.24 &  5.4\\
NGC 1187 &   37.9 &    4.7 &   74.1 &   16.1 &   1.96 &   0.59 &  5.1\\
NGC 1241 &   30.8 &    5.2 &   41.5 &    4.0 &   1.35 &   0.31 &  3.1\\
NGC 1302 &   30.7 &    4.0 &   50.4 &    4.6 &   1.64 &   0.32 &  0.0\\
NGC 1317 &   61.1 &    7.8 &   54.5 &   10.4 &   0.89 &   0.25 &  0.7\\
NGC 1832 &   18.9 &    3.3 &   32.8 &    3.1 &   1.74 &   0.40 &  4.1\\
NGC 3261 &   28.3 &    3.9 &   44.1 &    4.4 &   1.56 &   0.32 &  3.3\\
NGC 3275 &   28.9 &    4.4 &   44.2 &    4.6 &   1.53 &   0.34 &  1.8\\
NGC 3504 &   37.4 &    5.0 &   44.5 &    5.6 &   1.19 &   0.27 &  2.1\\
NGC 3507 &   28.1 &    4.0 &   34.2 &    2.6 &   1.21 &   0.23 &  3.0\\
NGC 3513 &   29.0 &    2.3 &   43.5 &    5.4 &   1.50 &   0.29 &  5.1\\
NGC 3583 &   25.9 &    4.0 &   32.1 &    1.8 &   1.24 &   0.23 &  3.2\\
NGC 3686 &   23.2 &    3.3 &   35.6 &    5.1 &   1.53 &   0.38 &  4.1\\
NGC 3726 &   42.8 &    6.8 &   83.5 &   13.9 &   1.95 &   0.55 &  5.2\\
NGC 4051 &   54.1 &    7.3 &   98.0 &   14.0 &   1.81 &   0.44 &  4.2\\
NGC 4123 &   59.2 &    8.2 &   69.1 &    9.4 &   1.17 &   0.28 &  4.8\\
NGC 4303 &   52.5 &   12.0 &   89.1 &    8.5 &   1.70 &   0.45 &  4.1\\
NGC 4314 &   82.9 &    7.1 &   81.7 &   10.2 &   0.99 &   0.19 &  1.1\\
NGC 4394 &   45.0 &    4.5 &   76.7 &   10.0 &   1.71 &   0.36 &  2.7\\
NGC 4450 &   49.7 &    6.8 &   52.3 &    4.4 &   1.05 &   0.20 &  2.0\\
NGC 4457 &   40.7 &    8.4 &   39.9 &    2.8 &   0.98 &   0.23 &  0.3\\
NGC 4548 &   75.8 &    6.8 &   95.2 &   11.9 &   1.26 &   0.25 &  3.0\\
NGC 4579 &   48.8 &    5.2 &   71.1 &    8.4 &   1.46 &   0.30 &  2.8\\
NGC 4643 &   66.4 &    9.3 &   69.1 &    4.6 &   1.04 &   0.19 & -0.1\\
NGC 4665 &   62.5 &   10.7 &   55.2 &   11.8 &   0.88 &   0.29 & -0.1\\
NGC 4902 &   26.1 &    3.9 &   44.3 &    4.4 &   1.70 &   0.37 &  2.9\\
NGC 4930 &   47.2 &    6.1 &   46.6 &    3.9 &   0.99 &   0.19 &  3.2\\
NGC 4995 &   30.8 &    3.8 &   64.2 &    5.5 &   2.09 &   0.39 &  3.1\\
NGC 5701 &   45.6 &    3.6 &   64.6 &   10.8 &   1.42 &   0.32 & -0.1\\
NGC 5850 &   75.8 &    9.8 &  105.1 &   11.7 &   1.39 &   0.29 &  3.0\\
NGC 5921 &   57.0 &    6.0 &   71.4 &    7.1 &   1.25 &   0.23 &  3.9\\
NGC 6384 &   31.8 &    7.5 &   72.5 &    8.1 &   2.28 &   0.64 &  3.8\\
NGC 6782 &   29.6 &    4.1 &   37.1 &    2.6 &   1.25 &   0.23 &  0.7\\
NGC 7552 &   65.7 &   12.4 &   65.0 &    5.9 &   0.99 &   0.23 &  2.2\\
NGC 7723 &   25.5 &    3.0 &   33.5 &    4.2 &   1.31 &   0.29 &  3.1\\

\hline
\end{tabular}
\end{minipage}
\end{table*}

\subsection{Comparison with previous pattern speed estimates}

Before going to analysing the results of the sample we first compare
our pattern speed determinations with previous estimates for these
galaxies. 
From literature we found published pattern speed estimates
for eight galaxies in our sample. Unfortunately, none of these was
based on direct TW-method, instead either morphological arguments or
simulation modelling was used to determine $\mathcal{R}$ (for NGC
1073 and NGC 4123 we found both morphological and model based
estimates of pattern speed). 

In three cases, NGC 1073, NGC 4123, and
NGC 5921, $R_{CR}$ was estimated by the method of
\citet{puerari97} \citep{aguerri98}: the $B$- and $I$-band $m=2$ phase
angle crossing was taken to be signature of CR, giving $\mathcal{R}=$
1.17, 1.41 and 1.28, respectively. These compare quite well with our
estimates ($\mathcal{R} = 0.95 \pm 0.24, 1.17 \pm 0.28$ and $1.25 \pm 0.23$).

For NGC 613 \citep{elmegreen92a} and NGC 3504 \citep{kenney93} the
morphological signatures of the outer Lindblad resonance (extent of
spiral structure, outer ring) and rotation curves were used to
determine $R_{CR}$ (in both cases corresponding to $\mathcal{R} =
1.0$). For NGC 7723 \citet{aguerri2000} determined $\mathcal{R}=1.0$
by assuming that the corotation region lack recent star formation. Of
these, the cases of NGC 3504 and NGC 7723 are in good agreement with
our value ($\mathcal{R} =1.19 \pm 0.27$, $\mathcal{R}=1.31 \pm 0.29$),
but for NGC 613 the difference is large (our value $\mathcal{R}=1.61
\pm 0.24$). However, our model of NGC 613 was not particularly good:
the fit is based just on the inner pseudoring and the innermost part
of the spiral arms. Perhaps the complicated spiral structure in beyond
our simple model.

Three galaxies of the sample have been previously
modelled. For NGC 1073 \citet{england90} found $\mathcal{R}=1.1$ (our
value $\mathcal{R}=0.95 \pm 0.24$), for
NGC 4123 \citet{weiner2001b} found $\mathcal{R}=1.35$ (our value
$\mathcal{R}=1.18 \pm 0.26$) and
for NGC 5850 \citet{aguerri2001} found $\mathcal{R}=1.35$ (our value
$\mathcal{R}=1.39 \pm 0.29$). In addition
to these, the pattern speed of NGC 4314
was determined by matching the shapes of particle orbits with the observed
outer ring \citep{quillen94} giving $\mathcal{R}=1.17$ (our value
$\mathcal{R}=0.99 \pm 0.19$). 

Altogether, in half of the eight cases a smaller value of
$\mathcal{R}$ was found than in our study, and in another half a
larger one was reported in the literature. For only one galaxy the
difference was clearly larger than our error estimate for
$\mathcal{R}$. If we take into account differences in adopted bar
lengths and orientation parameters, our models can be considered to be
consistent with previous pattern speed estimates for these galaxies.

\subsection{Dependence on Hubble type}

\begin{figure}
\resizebox{\hsize}{!}{\includegraphics{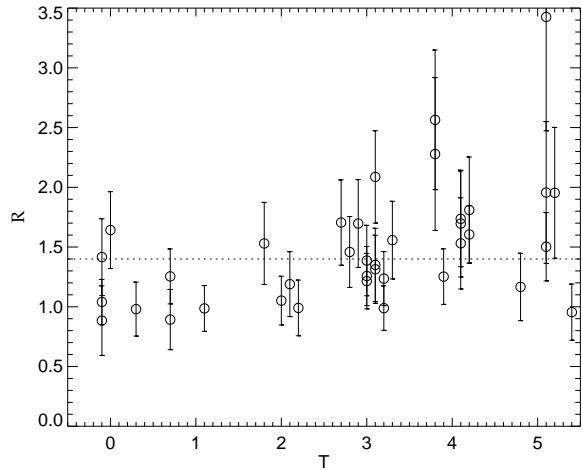}}
\caption{The best fitting $\mathcal{R}$ vs. Hubble stage $T$ (decimal
  number values from Hyperleda are used in this
  and other figures to increase the clarity of the plot by spreading
  the distribution data points in horizontal direction). Dashed line
  shows the upper limit for fast bars.}  
\label{coro_trc}
\end{figure}

Fig.~\ref{coro_trc} shows the determined values of $\mathcal{R}$ as a
function of Hubble stage $T$, taken from HyperLeda
\citep{paturel2003}. There seems to be a trend between the pattern
speed and the Hubble stage $T$: the average value of $\mathcal{R}$ and
its standard deviation are $1.15 \pm 0.25$ for SB0/a to SBab
($T=0-2$), $1.44 \pm 0.29$ for SBb ($T=3$) and $1.82 \pm 0.63$ for
SBbc to SBc ($T=4-5$). Thus, early type (SBab or earlier) galaxies are
typically fast rotators (in three cases the nomimal value $\mathcal{R}
> 1.4$, but the fast bar is within the error estimates), whereas later
type galaxies include both fast (10 cases) and slow rotators (16
cases). Especially, with five galaxies the fast rotating bar is ruled
out based on the adopted error estimates for the corotation resonance
radius and bar size.

\begin{figure}
\resizebox{\hsize}{!}{\includegraphics{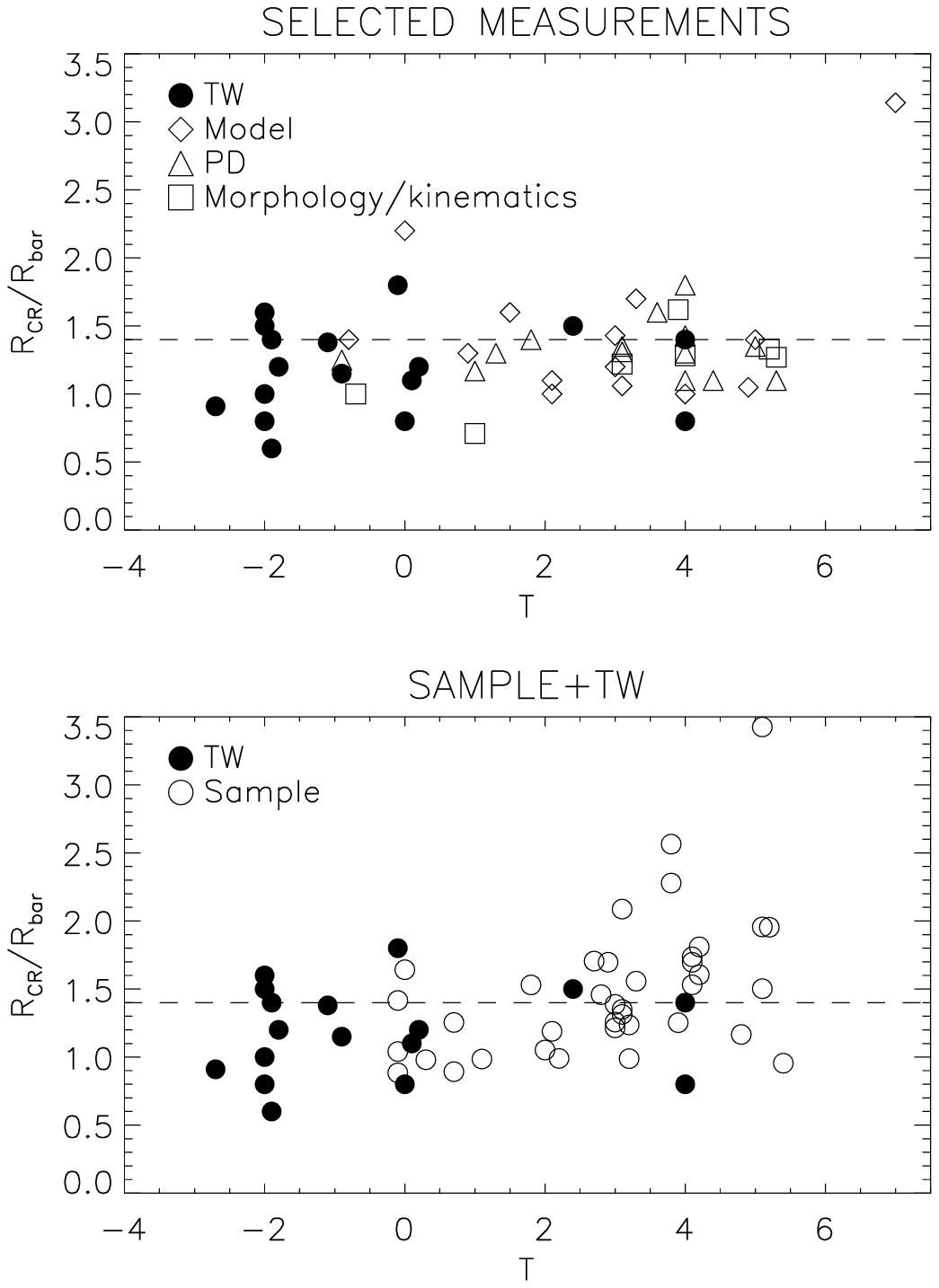}}
\caption{Top: a comparison sample of previous determinations of bar
  pattern speeds as a function of $\mathcal{R}$. Filled circles show
  pattern speed estimates based on Tremaine-Weinberg method, triangles
  estimates based on modelling, squares estimates based on
  Puerari-Dottori method and diamonds estimates based on morphological
  or kinematical arguments. Bottom: The results of our sample compared with
  results obtained with Tremaine-Weinberg method.} 
\label{otos_compared}
\end{figure}

Our results can be compared with other pattern speed estimates also in
the sample level -- did we find a similar trend as the other
researchers. For this purpose we have composed a comparison sample of
pattern speed determinations from literature by the following
criteria: 1) using the TW-method, 2) using simulation modelling, 3)
using the Puerari--Dottori -method, and 4) galaxies whose pattern
speed estimate is based on morphological arguments. Each galaxy enters
the sample only once: if its pattern speed is estimated by TW-method,
then its pattern speed estimates by other methods are omitted and so
on. If essentially the same method is used several times for a galaxy,
then the newest measurement is selected. The comparison sample is
listed in Table 2 and the comparison with our modelling is illustrated in
Fig.~\ref{otos_compared}.

\begin{table*}
\begin{minipage}{126mm}
\caption{Comparison sample.}
\begin{tabular}{lccccc}
\hline
Galaxy	&	T	&$\mathcal{R}$	 & Reference & Comments\\
TW-measurements&&&\\
\hline
E 139-09&	-2.0&	0.80&		\citet{aguerri2003} & \\	
E 281-31&	-0.1&	1.8&		\citet{gerssen2003}	& \\
IC 874& 	-1.9&	1.40&		\citet{aguerri2003}	&\\	
NGC 271&	2.4&	1.5&		\citet{gerssen2003}	&\\	
NGC 936&	-1.1&	1.38&
\citet{merrifield95} & see also \citet{kent89}	\\
NGC 1023&	-2.7&	0.91&		\citet{debattista2002}	&
higher $\mathcal{R}$ of two solutions\\
NGC 1308&	0.0&	0.80&		\citet{aguerri2003}	&	\\
NGC 1358&	0.2&	1.2&		\citet{gerssen2003}	&	\\
NGC 1440&	-2.0&	1.60&		\citet{aguerri2003}	&	\\
NGC 2523&	4.0&	1.4&	        \citet{treuthardt2007}	&	\\
NGC 2950&	-2.0&	1.0&		\citet{corsini2003}&	\\	
NGC 3412&	-2.0&	1.50&		\citet{aguerri2003}	&	\\
NGC 3992&	4.0&	0.80&		\citet{gerssen2003}	&	\\
NGC 4245&	0.1&	1.1&	        \citet{treuthardt2007}	&	\\
NGC 4431&	-1.9&	0.6&	        \citet{corsini2007} 	&	\\
NGC 4596&	-0.9&	1.15&		\citet{gerssen99} &	\\
NGC 7079&	-1.8&	1.2&		\citet{debattista2004}&	\\
\hline
Modelling & & &\\
\hline
ESO 566-24&	3.6&	1.60&           \citet{rautiainen2004}&	\\
IC 4214	&       1.8&	1.4&            \citet{salo99}	&	\\
NGC 157&	4.0&	1.43&           \citet{sempere97b}	&	\\
NGC 1073&	5.3&	1.10&		\citet{england90}  &	\\
NGC 1300&	4.0&	1.30&           \citet{lindblad96b}	&	alternate fit gives $\mathcal{R}=2.4$\\
NGC 1365&	3.1&	1.31 &	        \citet{lindblad96a} &  \\
NGC 1433&	1.5 &   1.60&		\citet{treuthardt2008}	&\\ 			
NGC 2336&	4.0&	1.10&	        \citet{wilke99}	&	\\
NGC 3783&	1.3&	1.3&		\citet{garbar99} &	\\
NGC 4123&	5.0&	1.35&		\citet{weiner2001b} &	\\
NGC 4314&	1.0&	1.17&		\citet{quillen94} &	orbits in determined potential\\
NGC 4321&	4.0&	1.80&		\citet{garbur98}	&	\\
NGC 5850&	3.1&	1.35&		\citet{aguerri2001}	&	\\
NGC 7479&	4.4&	1.1&	         \citet{wilke2000}	&	\\
\hline
Puerari\&Dottori method & & &\\
\hline
NGC 1530 &	3.1&	1.22&		\citet{aguerri98}	&\\	
NGC 2273&	1.0&	1.27&		\citet{aguerri98}	&\\		
NGC 3516&	-2.0&	0.71&		\citet{aguerri98}	&\\		
NGC 3359&	5.2&	1.33&		\citet{aguerri98}	&\\		
NGC 5921&	4.0&	1.28&		\citet{aguerri98}	&\\		
NGC 6951&	3.9&	1.62&		\citet{aguerri98}	&\\		
NGC 7743&	-0.7&	1.0&		\citet{aguerri98}	&\\		
\hline
Morphological or \\
kinematical arguments & & &\\
\hline
ESO 509-98&	0.9&    1.30&	        \citet{buta98b}   &\\ 		
IC 4290	&       3.3&    1.70&		\citet{buta98b}	&\\ 			
NGC 0613&	4.0&    1.00&		\citet{elmegreen92a}	&\\ 		
NGC 0925&	7.0&    3.14&           \citet{elmegreen98c}     & kinematical identification or CR\\ 
NGC 1326&	-0.8&   1.40&		\citet{buta98b}	&\\ 			
NGC 3081&	0.0 &   2.20&		\citet{buta98a}	&\\ 			
NGC 3351&	3.0 &   1.20&		\citet{devereux92}	&\\ 			
NGC 3504&	2.1 &   1.00&		\citet{kenney93}	&\\ 			
NGC 3627&	3.0 &   1.43&		\citet{chemin2003}	&\\ 			
NGC 4151&	2.1 &   1.1&		\citet{mundell99b}	&\\ 			
NGC 5236&	5.0 &   1.40&		\citet{kenney91}	&\\ 			
NGC 6221&	4.9 &   1.05&		\citet{vega98}	&\\ 		
NGC 7723&	3.1 &   1.06 &		\citet{aguerri2000}		&\\ 		
\hline
\end{tabular}
\end{minipage}
\end{table*}

Strictly speaking, our results are in
agreement with those obtained with TW-method. Most of the
measurements with TW-method have been limited to early type
galaxies, which indeed have fast bars. This has been taken to indicate
that the dark halo contribution in these galaxies is so low that it
cannot decelerate the bar rotation
\citep{debattista2000,aguerri2003}. If so, then the slow
bars in our sample could be systems where angular
momentum transfer between the bar and the dark halo with a substantial
central density has slowed down bar rotation. Alternatively, the
difference in pattern speeds can be primordial: the early type
galaxies, which have massive bulges, may favor the formation of fast bars
\citep{combes93}. 

In future it would be important to increase the overlap of pattern speed
estimates based on different methods both by extending the current
modelling to earlier type galaxies \citep[e.g.\ by using the data of
  the ongoing survey of S0-galaxies, ][]{laurikainen2005,buta2006a}
and by extending the direct measurements to later types. This would
give a better understanding of accuracy and possible bias of pattern
speed determinations done with different methods.

The slow bars in our sample cannot be refuted by the ambiguity in the
determination of bar length: our values tend to be slightly {\it larger} than
previous bar length estimates for the same galaxies, thus favoring a
faster bar. Furthermore, the comparison with other estimates of
corotation radius for the sample galaxies does not show any systematic
difference. If our error estimates for $R_{CR}$ and $R_{bar}$ (as
listed in Table 1) hold, then of 19 galaxies with nominal value
$\mathcal{R} > 1.4$, we can coin five (six if we adopt the alternative
pattern speed solution for NGC 7723) as ``definitely slow rotating'',
i.e. $\mathcal{R} - \Delta \mathcal{R} > 1.4$ (see
Fig.~\ref{coro_trc}), and several others for which a fast bar solution
in unlikely. Of these, NGC 289 and NGC 578 both have a very small bar
when compared to the extent of the spiral structure. The innermost
structure of these galaxies would allow also higher pattern speed than
the best fitting model, but even then $\mathcal{R}$ would be much
larger than 1.4. In NGC 1832 the best-fitting model, which gives a
slow bar, reproduces the inner ring. The ring disappears when the
pattern speed is increased to fast bar domain. The bar does not fill
the inner ring, thus also the morphology supports slow bar
solution. The innermost structure of NGC 3726 allows also a faster bar
pattern speed, but even the inner ring morphology is better reproduced
with the value listed in Table 1, which gives a good
overall fit. The straight arm segments of the inner spiral structure
in NGC 4303 are well reproduced with the listed value of
$\mathcal{R}$, but disappear when reaching the fast bar
domain. However, the bar of this galaxy has leading offset dust lanes,
which are usually taken as a signature of fast-rotating bar. Our gas
dynamical model is not well-suited for modelling the dust lanes, but
indeed we got a better fit for these features with fast-rotating
bar. However, the spiral response with such pattern speed choice is
quite bad. In the case of NGC 4995 the fit is not very good and a
value giving a fast rotating bar cannot be excluded. NGC 6384 has very
long spiral arms, which are completely reproduced with the listed
value of $\mathcal{R}$. Only the innermost part can be reproduced with
a fast rotating bar.

An alternative explanation to slow bars, that the spiral structure
in late type galaxies typically rotates with a lower pattern speed
than the bar, and that our method gives too much emphasis to the spiral,
is certainly possible for some galaxies of the sample (NGC 3726, NGC
4303, NGC 4995, NGC 6384). However, there are several
galaxies (NGC 289, NGC 578, NGC 1832), where also the fit to the
innermost features favours low pattern speed. If the bar in these
galaxies is fast rotating, then even the innermost part of the spiral
rotates with a lower pattern speed, and the bar may be relatively
insignificant in their dynamics. 

\subsection{Size and strength of the bar}

\begin{figure}
\resizebox{\hsize}{!}{\includegraphics{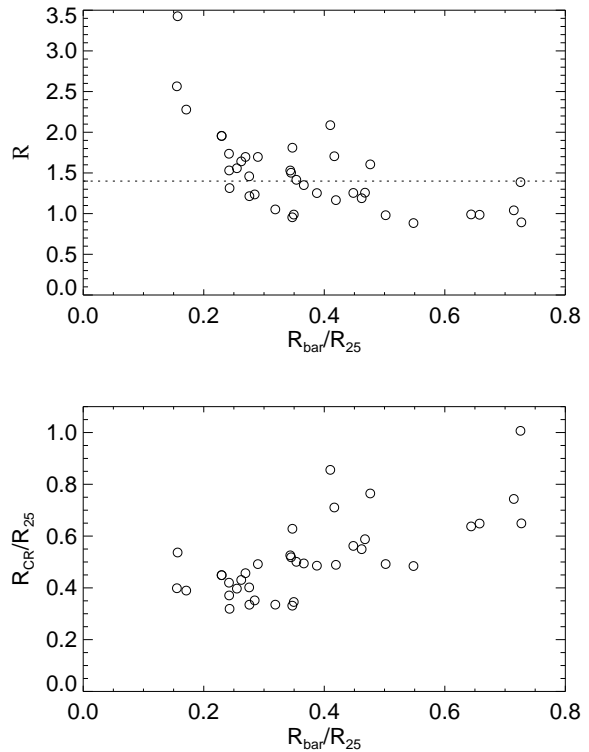}}
\caption{Top: best fitting $\mathcal{R}$ vs. relative size of the bar
  $R_{bar}/R_{25}$, bottom: relative corotation radius $R_{CR}/R_{25}$
  vs.$R_{bar}/R_{25}$.} 
\label{coro_barsize}
\end{figure}

The comparison of the determined corotation radius to the galaxy size,
the latter estimated to be proportional to isophotal radius $R_{25}$, does not
show any clear trend with galaxy morphology. Regardless of the Hubble
type, $R_{CR}$ can be anything between 0.3 and 1.0 times $R_{25}$. The
situation is different if we compare the dimensionless pattern speed
$\mathcal{R}$ to relative bar size, measured by the estimated semi
major axes $R_{bar}$, and scaled to the isophotal radius $R_{25}$
(Fig.~\ref{coro_barsize}, top. Now there is a clear correlation: the
slow bars tend to be also the smallest ones: with one exception, the galaxies with
$R_{bar}/R_{25} \le 0.25$ have $\mathcal{R} \ge 1.4$.  When all the
galaxies are plotted in $R_{CR}/R_{25}$ vs. $R_{bar}/R_{25}$ one can
see for the smallest bars ($R_{bar}/R_{25} \approx 0.15 - 0.35$), the
corotation resonance is close to 0.4 $R_{25}$. This
corresponds to typical values for the radii, where
\citet{elmegreen95a} found the end of the symmetric two-armed spiral to be
in non-barred galaxies and galaxies with small bars.
For the larger bars, the bar size essentially
correlates with the corotation radius (Fig.~\ref{coro_barsize},
bottom), but the scatter is large. 

In some studies \citep{patsis91,kranz2003} it has been suggested that
in non-barred galaxies the end of the spiral or at least
the end of its symmetric part lies near corotation or
inner 4/1-resonance. The possible independence of corotation resonance radius
from the bar size in the galaxies with smallest bars could suggest
that the spiral in these galaxies indeed has a different pattern
speed, whose corotation radius is regulated by the galaxy size.

\begin{figure}
\resizebox{\hsize}{!}{\includegraphics{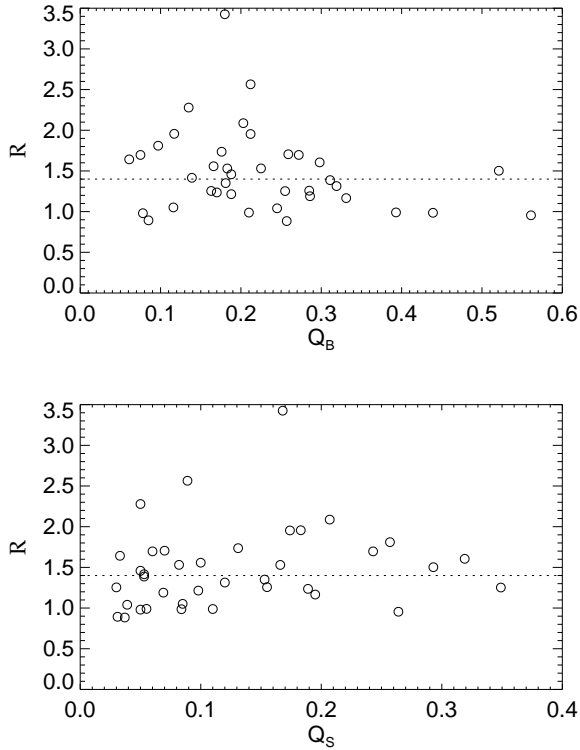}}
\caption{Top: best fitting $\mathcal{R}$ vs. bar strength
  $Q_B$, bottom:
  $\mathcal{R}$ vs. spiral strength $Q_S$.} 
\label{coro_barstrength}
\end{figure}

In Fig.~\ref{coro_barstrength} we compare $\mathcal{R}$ to parameters
$Q_B$ and $Q_S$ from \citet{buta2005}, which characterize the
strengths of bar and spiral components (maximum tangential force
divided by the azimuthally averaged radial force in each radius). With
one exception, galaxies with slow bars ($\mathcal{R} > 1.4$) have $Q_B
\le 0.3$. However, there are several galaxies with similar bar
strength but fast bar. The combined results with bar size and strength
are in accordance with \citet{elmegreen95a} who found by morphological
arguments that the flat (and typically long) bars tend to be faster
rotators than the exponential (and typically short) bars. There is no
clear correlation between pattern speed and spiral strength.

\subsection{Bulge-to-total flux ratio, galaxy luminosity and color}

\begin{figure}
\resizebox{\hsize}{!}{\includegraphics{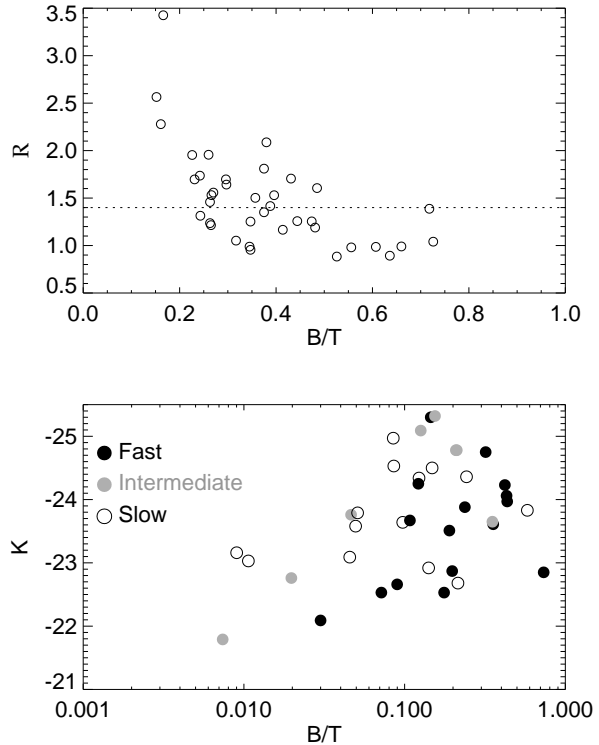}}
\caption{Correlation with bulge flux fraction. Top: $\mathcal{R}$
  vs. $B/T$, bottom: fast (black circles), slow (open circles) and
  intermediate (grey circles) bars in $(B/T,K)$-plane.}
\label{coro_bt_etc}
\end{figure}

The top frame of Fig.~\ref{coro_bt_etc} shows $\mathcal{R}$ as a
function of bulge-to-total flux fraction ($B/T$-ratio) in $H$-band
\citep[see also][]{salo2007}. The apparent bulge contribution is one
of the criteria in Hubble classification, so it is not surprising that
we get almost a mirror image of Fig.~\ref{coro_trc}.

We studied the dependency between $\mathcal{R}$ and absolute
$B$-luminosity (calculated from the $B$-magnitudes in RC3) and distances by
\citet{tully88}, but found no clear correlation. The same was the
situation with absolute $K$-magnitudes based on
2MASS data \citep{jarrett2003} and neither there was correlation with
colour. However, if we plot the galaxies in $(B/T,K)$-plane
(Fig.~\ref{coro_bt_etc}, bottom), then the fast and slow galaxies
roughly occupy different areas: slower bar favouring lower bulge
fraction and/or higher galaxy luminosity.   

\subsection{Pattern speed vs. dust lane morphology}

According to gas dynamical simulations by \citet{athanassoula92b}, the
existence of leading offset dust lanes in the bars tells us two
things: that there is an ILR and that $\mathcal{R}=1.2 \pm 0.2$. The
first condition comes from the existence of so-called $x_2$-orbits,
which require the presence of an ILR: the
dust lanes form as the orientation of the orbits of gas clouds change
from parallel orientation of $x_1$-orbits (with respect to the bar) in
the outer parts of the bar to the perpendicular orientation of
$x_2$-orbits in the inner parts. The second condition comes from the
gas dynamical simulation, only in this parameter region the shape of
the shock corresponds to the observed dust morphology. 

We have checked the dust lane morphology of the sample galaxies from
OSUBSGS images and images from various other sources, e.g.\ NED and
the web page of de Vaucouleurs Atlas of Galaxies
\citep{buta2007a}. Omitting cases where the inclination makes
recognition and classification of dust lanes ambiguous, we found that
more than third of the galaxies in the sample does not show classic
dust lane morphology. In some of the cases the dust lanes are missing
altogether, perhaps due to lack of interstellar matter in the bar
region, in others the morphology is more complicated or even
chaotic. Of the seven galaxies with $\mathcal{R}_{min} > 1.4$, the
dust lane morphology is uncertain due to inclination for NGC 3726 and
NGC 4995. In NGC 289 the dust lane morphology is not of the
``classic'' type, in NGC 578 it is more or less chaotic. NGC 1832 and
NGC 6384 do not show dust lanes in the bar region. In NGC 4303 the
dust lane morphology is of classic type suggesting that the bar in
this galaxies may be fast rotating. Indeed, if we compare a model for
NGC 4303 with $\mathcal{R} \approx 1$ (Fig.~\ref{ngc4303_pspeeds}),
then the gas morphology in the bar region shows features that look a
bit like the observed dust lanes, but the spiral structure is not well
reproduced.

\section{Conclusions}

We have modelled 38 barred galaxies with simulations using
collisionless test particles and inelastatically colliding test
particles. The gravitational potentials used were derived from
$H$-band images from Ohio State University Bright Spiral Galaxy Survey.

Our findings are as follows:

1) the average value of the dimensionless pattern speed,
   $\mathcal{R}=R_{CR}/R_{bar}$ depends on the morphological type,
   being about 1.2 for the early type barred galaxies (SB0/a--SBab), 1.4 to
   intermediate type (SBb) and 1.8 to late type (SBbc--SBc). 

2) When error estimates are considered, all early type spiral galaxies in the
   sample ($T \le 2$) are in accordance with having fast bars
   ($\mathcal{R} < 1.4$). 

3) If the derived pattern speed corresponds to that of the bar, then
intermediate and late type galaxies include both fast and slow
bars. In five cases the bar remains in slow bar domain even if a
``worst case scenario'' allowed by the error estimates in pattern
speed and bar size determinations is considered. However, the
existence of multiple pattern speeds, i.e.\ spiral rotating more
slowly than the bar, is a viable alternative in some cases, especially
for NGC 4303.

4) Slow bars are short when compared to isophotal radius
$R_{25}$. Especially, with only one exception, galaxies with
$R_{bar}/R_{25} \le 0.25$ have a slow bar.

5) With one exception, the slow bars have bar strength $Q_B \le 0.3$,
although several bars with similar strength are also found to be fast.

6) There is no clear correlation with the absolute $B$- or $K$-magnitudes
   of the galaxies or the $B-K$ color index.

7) Galaxies with fast and slow bars seem to occupy different regions
when plotted in a two-dimensional coordinate system defined by
bulge-to-total flux ratio and absolute $K$-magnitude.

8) Omitting possible systematic errors, the error estimates of the
   model-based pattern speeds are typically
   smaller than with Tremaine-Weinberg method. If the results by these
   two methods are found to be consistent, then modelling would enable
   pattern speed estimation for large galaxy samples.

\section*{Acknowledgments}

We thank the referee, whose suggestions helped us to improve the manuscript.

The financial support from the Magnus Ehrnrooth foundation and the
Academy of Finland is acknowledged.

This work made use of data from the Ohio State University Bright
Spiral Galaxy Survey, which was funded by grants AST-9217716 and
AST-9617006 from the United States National Science Foundation, with
additional support from the Ohio State University. 

We acknowledge the usage of the HyperLeda database
(http://leda.univ-lyon1.fr).  

We have also used the NASA/IPAC Extragalactic Database
(NED) which is operated by the Jet Propulsion Laboratory, California
Institute of Technology, under contract with the National Aeronautics
and Space Administration. 

\bibliographystyle{mn2e}
\bibliography{astrobib}

\begin{thebibliography}{}

\bibitem[\protect\citeauthoryear{{Abraham}, {Merrifield}, {Ellis}, {Tanvir} \&
  {Brinchmann}}{{Abraham} et~al.}{1999}]{abraham99}
{Abraham} R.~G.,  {Merrifield} M.~R.,  {Ellis} R.~S.,  {Tanvir} N.~R.,
  {Brinchmann} J.,  1999, MNRAS, 308, 569

\bibitem[\protect\citeauthoryear{{Aguerri}, {Beckman} \& {Prieto}}{{Aguerri}
  et~al.}{1998}]{aguerri98}
{Aguerri} J. A.~L.,  {Beckman} J.~E.,    {Prieto} M.,  1998, AJ, 116, 2136

\bibitem[\protect\citeauthoryear{{Aguerri}, {Debattista} \&
  {Corsini}}{{Aguerri} et~al.}{2003}]{aguerri2003}
{Aguerri} J. A.~L.,  {Debattista} V.~P.,    {Corsini} E.~M.,  2003, MNRAS, 338,
  465

\bibitem[\protect\citeauthoryear{{Aguerri}, {Hunter}, {Prieto}, {Varela},
  {Gottesman} \& {Mu{\~ n}oz-Tu{\~ n}{\' o}n}}{{Aguerri}
  et~al.}{2001}]{aguerri2001}
{Aguerri} J.~A.~L.,  {Hunter} J.~H.,  {Prieto} M.,  {Varela} A.~M.,
  {Gottesman} S.~T.,    {Mu{\~ n}oz-Tu{\~ n}{\' o}n} C.,  2001, A\&A, 373, 786

\bibitem[\protect\citeauthoryear{{Aguerri}, {Mu{\~n}oz-Tu{\~n}{\'o}n}, {Varela}
  \& {Prieto}}{{Aguerri} et~al.}{2000}]{aguerri2000}
{Aguerri} J. A.~L.,  {Mu{\~n}oz-Tu{\~n}{\'o}n} C.,  {Varela} A.~M.,    {Prieto}
  M.,  2000, A\&A, 361, 841

\bibitem[\protect\citeauthoryear{{Athanassoula}}{{Athanassoula}}{1992}]{athana%
ssoula92b}
{Athanassoula} E.,  1992, MNRAS, 259, 345

\bibitem[\protect\citeauthoryear{{Athanassoula}}{{Athanassoula}}{2003}]{athana%
ssoula2003}
{Athanassoula} E.,  2003, MNRAS, 341, 1179

\bibitem[\protect\citeauthoryear{{Athanassoula} \& {Misiriotis}}{{Athanassoula}
  \& {Misiriotis}}{2002}]{athanassoula2002a}
{Athanassoula} E.,  {Misiriotis} A.,  2002, MNRAS, 330, 35

\bibitem[\protect\citeauthoryear{{Bournaud}, {Combes} \& {Semelin}}{{Bournaud}
  et~al.}{2005}]{bournaud2005}
{Bournaud} F.,  {Combes} F.,    {Semelin} B.,  2005, MNRAS, 364, L18

\bibitem[\protect\citeauthoryear{{Buta}, {Alpert}, {Cobb}, {Crocker} \&
  {Purcell}}{{Buta} et~al.}{1998}]{buta98b}
{Buta} R.,  {Alpert} A.~J.,  {Cobb} M.~L.,  {Crocker} D.~A.,    {Purcell}
  G.~B.,  1998, AJ, 116, 1142

\bibitem[\protect\citeauthoryear{{Buta} \& {Block}}{{Buta} \&
  {Block}}{2001}]{buta2001b}
{Buta} R.,  {Block} D.~L.,  2001, ApJ, 550, 243

\bibitem[\protect\citeauthoryear{{Buta} \& {Combes}}{{Buta} \&
  {Combes}}{1996}]{buta96b}
{Buta} R.,  {Combes} F.,  1996, Fundamentals of Cosmic Physics, 17, 95

\bibitem[\protect\citeauthoryear{{Buta}, {Corwin} \& {Odewahn}}{{Buta}
  et~al.}{2007}]{buta2007a}
{Buta} R.,  {Corwin} J.,    {Odewahn} S.,  2007, The de Vaucouleurs Atlas of
  Galaxies.
New York: Cambridge University Press

\bibitem[\protect\citeauthoryear{{Buta}, {Laurikainen} \& {Salo}}{{Buta}
  et~al.}{2004}]{buta2004}
{Buta} R.,  {Laurikainen} E.,    {Salo} H.,  2004, AJ, 127, 279

\bibitem[\protect\citeauthoryear{{Buta}, {Laurikainen}, {Salo}, {Block} \&
  {Knapen}}{{Buta} et~al.}{2006}]{buta2006a}
{Buta} R.,  {Laurikainen} E.,  {Salo} H.,  {Block} D.~L.,    {Knapen} J.~H.,
  2006, AJ, 132, 1859

\bibitem[\protect\citeauthoryear{{Buta} \& {Purcell}}{{Buta} \&
  {Purcell}}{1998}]{buta98a}
{Buta} R.,  {Purcell} G.~B.,  1998, AJ, 115, 484

\bibitem[\protect\citeauthoryear{{Buta}, {Vasylyev}, {Salo} \&
  {Laurikainen}}{{Buta} et~al.}{2005}]{buta2005}
{Buta} R.,  {Vasylyev} S.,  {Salo} H.,    {Laurikainen} E.,  2005, AJ, 130, 506

\bibitem[\protect\citeauthoryear{{Byrd}, {Rautiainen}, {Salo}, {Buta} \&
  {Crocher}}{{Byrd} et~al.}{1994}]{byrd94}
{Byrd} G.,  {Rautiainen} P.,  {Salo} H.,  {Buta} R.,    {Crocher} D.~A.,  1994,
  AJ, 108, 476

\bibitem[\protect\citeauthoryear{{Canzian}}{{Canzian}}{1993}]{canzian93}
{Canzian} B.,  1993, ApJ, 414, 487

\bibitem[\protect\citeauthoryear{{Chemin}, {Cayatte}, {Balkowski}, {Marcelin},
  {Amram}, {van Driel} \& {Flores}}{{Chemin} et~al.}{2003}]{chemin2003}
{Chemin} L.,  {Cayatte} V.,  {Balkowski} C.,  {Marcelin} M.,  {Amram} P.,  {van
  Driel} W.,    {Flores} H.,  2003, A\&A, 405, 89

\bibitem[\protect\citeauthoryear{{Chernin}, {Zasov}, {Arkhipova} \&
  {Kravtsova}}{{Chernin} et~al.}{2000}]{chernin2000}
{Chernin} A.~D.,  {Zasov} A.~V.,  {Arkhipova} V.~P.,    {Kravtsova} A.~S.,
  2000, Astronomy Letters, 26, 285

\bibitem[\protect\citeauthoryear{{Combes} \& {Elmegreen}}{{Combes} \&
  {Elmegreen}}{1993}]{combes93}
{Combes} F.,  {Elmegreen} B.~G.,  1993, A\&A, 271, 391

\bibitem[\protect\citeauthoryear{{Contopoulos}}{{Contopoulos}}{1980}]{contopou%
los80a}
{Contopoulos} G.,  1980, A\&A, 81, 198

\bibitem[\protect\citeauthoryear{{Corsini}, {Aguerri}, {Debattista},
  {Pizzella}, {Barazza} \& {Jerjen}}{{Corsini} et~al.}{2007}]{corsini2007}
{Corsini} E.~M.,  {Aguerri} J.~A.~L.,  {Debattista} V.~P.,  {Pizzella} A.,
  {Barazza} F.~D.,    {Jerjen} H.,  2007, ApJ, 659, L121

\bibitem[\protect\citeauthoryear{{Corsini}, {Debattista} \&
  {Aguerri}}{{Corsini} et~al.}{2003}]{corsini2003}
{Corsini} E.~M.,  {Debattista} V.~P.,    {Aguerri} J.~A.~L.,  2003, ApJ, 599,
  L29

\bibitem[\protect\citeauthoryear{{de Grijs}}{{de Grijs}}{1998}]{degrijs98}
{de Grijs} R.,  1998, MNRAS, 299, 595

\bibitem[\protect\citeauthoryear{{de Vaucouleurs}, {de Vaucouleurs}, {Corwin},
  {Buta}, {Paturel} \& {Fouque}}{{de Vaucouleurs}
  et~al.}{1991}]{devaucouleurs91}
{de Vaucouleurs} G.,  {de Vaucouleurs} A.,  {Corwin} H.~G.,  {Buta} R.~J.,
  {Paturel} G.,    {Fouque} P.,  1991, {Third Reference Catalogue of Bright
  Galaxies}.
~ Springer-Verlag Berlin Heidelberg New York

\bibitem[\protect\citeauthoryear{{Debattista}}{{Debattista}}{2003}]{debattista%
2003}
{Debattista} V.~P.,  2003, MNRAS, 342, 1194

\bibitem[\protect\citeauthoryear{{Debattista}, {Corsini} \&
  {Aguerri}}{{Debattista} et~al.}{2002}]{debattista2002}
{Debattista} V.~P.,  {Corsini} E.~M.,    {Aguerri} J.~A.~L.,  2002, MNRAS, 332,
  65

\bibitem[\protect\citeauthoryear{{Debattista} \& {Sellwood}}{{Debattista} \&
  {Sellwood}}{1998}]{debattista98}
{Debattista} V.~P.,  {Sellwood} J.~A.,  1998, ApJ, 493, L5

\bibitem[\protect\citeauthoryear{{Debattista} \& {Sellwood}}{{Debattista} \&
  {Sellwood}}{2000}]{debattista2000}
{Debattista} V.~P.,  {Sellwood} J.~A.,  2000, ApJ, 543, 704

\bibitem[\protect\citeauthoryear{{Debattista} \& {Williams}}{{Debattista} \&
  {Williams}}{2004}]{debattista2004}
{Debattista} V.~P.,  {Williams} T.~B.,  2004, ApJ, 605, 714

\bibitem[\protect\citeauthoryear{{Devereux}, {Kenney} \& {Young}}{{Devereux}
  et~al.}{1992}]{devereux92}
{Devereux} N.~A.,  {Kenney} J.~D.,    {Young} J.~S.,  1992, AJ, 103, 784

\bibitem[\protect\citeauthoryear{{Elmegreen} \& {Elmegreen}}{{Elmegreen} \&
  {Elmegreen}}{1985}]{elmegreen85}
{Elmegreen} B.~G.,  {Elmegreen} D.~M.,  1985, ApJ, 288, 438

\bibitem[\protect\citeauthoryear{{Elmegreen}, {Elmegreen}, {Chromey},
  {Hasselbacher} \& {Bissell}}{{Elmegreen} et~al.}{1996a}]{elmegreen96b}
{Elmegreen} B.~G.,  {Elmegreen} D.~M.,  {Chromey} F.~R.,  {Hasselbacher} D.~A.,
     {Bissell} B.~A.,  1996a, AJ, 111, 2233

\bibitem[\protect\citeauthoryear{{Elmegreen}, {Elmegreen} \&
  {Hirst}}{{Elmegreen} et~al.}{2004}]{elmegreen2004}
{Elmegreen} B.~G.,  {Elmegreen} D.~M.,    {Hirst} A.~C.,  2004, ApJ, 612, 191

\bibitem[\protect\citeauthoryear{{Elmegreen}, {Elmegreen} \&
  {Montenegro}}{{Elmegreen} et~al.}{1992}]{elmegreen92a}
{Elmegreen} B.~G.,  {Elmegreen} D.~M.,    {Montenegro} L.,  1992, ApJS, 79, 37

\bibitem[\protect\citeauthoryear{{Elmegreen}, {Wilcots} \&
  {Pisano}}{{Elmegreen} et~al.}{1998}]{elmegreen98c}
{Elmegreen} B.~G.,  {Wilcots} E.,    {Pisano} D.~J.,  1998, ApJ, 494, L37

\bibitem[\protect\citeauthoryear{{Elmegreen}, {Bellin} \&
  {Elmegreen}}{{Elmegreen} et~al.}{1990}]{elmegreen90c}
{Elmegreen} D.~M.,  {Bellin} A.~D.,    {Elmegreen} B.~G.,  1990, ApJ, 364, 415

\bibitem[\protect\citeauthoryear{{Elmegreen} \& {Elmegreen}}{{Elmegreen} \&
  {Elmegreen}}{1995}]{elmegreen95a}
{Elmegreen} D.~M.,  {Elmegreen} B.~G.,  1995, ApJ, 445, 591

\bibitem[\protect\citeauthoryear{{Elmegreen}, {Elmegreen}, {Chromey},
  {Hasselbacher} \& {Bissell}}{{Elmegreen} et~al.}{1996b}]{elmegreen96a}
{Elmegreen} D.~M.,  {Elmegreen} B.~G.,  {Chromey} F.~R.,  {Hasselbacher} D.~A.,
     {Bissell} B.~A.,  1996b, AJ, 111, 1880

\bibitem[\protect\citeauthoryear{{England}, {Gottesman} \& {Hunter}}{{England}
  et~al.}{1990}]{england90}
{England} M.~N.,  {Gottesman} S.~T.,    {Hunter} J.~H.,  1990, ApJ, 348, 456

\bibitem[\protect\citeauthoryear{{Erwin}}{{Erwin}}{2004}]{erwin2004}
{Erwin} P.,  2004, A\&A, 415, 941

\bibitem[\protect\citeauthoryear{{Erwin}}{{Erwin}}{2005}]{erwin2005}
{Erwin} P.,  2005, MNRAS, 364, 283

\bibitem[\protect\citeauthoryear{{Eskridge}, {Frogel}, {Pogge}, {Quillen},
  {Berlind}, {Davies}, {DePoy}, {Gilbert}, {Houdashelt}, {Kuchinski}, {Ram{\'
  i}rez}, {Sellgren}, {Stutz}, {Terndrup} \& {Tiede}}{{Eskridge}
  et~al.}{2002}]{eskridge2002}
{Eskridge} P.~B.,  {Frogel} J.~A.,  {Pogge} R.~W.,  {Quillen} A.~C.,  {Berlind}
  A.~A.,  {Davies} R.~L.,  {DePoy} D.~L.,  {Gilbert} K.~M.,  {Houdashelt}
  M.~L.,  {Kuchinski} L.~E.,  {Ram{\' i}rez} S.~V.,  {Sellgren} K.,  {Stutz}
  A.,  {Terndrup} D.~M.,    {Tiede} G.~P.,  2002, ApJS, 143, 73

\bibitem[\protect\citeauthoryear{{Eskridge}, {Frogel}, {Pogge}, {Quillen},
  {Davies}, {DePoy}, {Houdashelt}, {Kuchinski}, {Ram{\'i}rez}, {Sellgren},
  {Terndrup} \& {Tiede}}{{Eskridge} et~al.}{2000}]{eskridge2000}
{Eskridge} P.~B.,  {Frogel} J.~A.,  {Pogge} R.~W.,  {Quillen} A.~C.,  {Davies}
  R.~L.,  {DePoy} D.~L.,  {Houdashelt} M.~L.,  {Kuchinski} L.~E.,
  {Ram{\'i}rez} S.~V.,  {Sellgren} K.,  {Terndrup} D.~M.,    {Tiede} G.~P.,
  2000, AJ, 119, 536

\bibitem[\protect\citeauthoryear{{Friedli} \& {Martinet}}{{Friedli} \&
  {Martinet}}{1993}]{friedli93b}
{Friedli} D.,  {Martinet} L.,  1993, A\&A, 277, 27

\bibitem[\protect\citeauthoryear{{Garc{\'{\i}}a-Barreto}, {Combes},
  {Koribalski} \& {Franco}}{{Garc{\'{\i}}a-Barreto} et~al.}{1999}]{garbar99}
{Garc{\'{\i}}a-Barreto} J.~A.,  {Combes} F.,  {Koribalski} B.,    {Franco} J.,
  1999, A\&A, 348, 685

\bibitem[\protect\citeauthoryear{{Garcia-Burillo}, {Sempere}, {Combes} \&
  {Neri}}{{Garcia-Burillo} et~al.}{1998}]{garbur98}
{Garcia-Burillo} S.,  {Sempere} M.~J.,  {Combes} F.,    {Neri} R.,  1998, A\&A,
  333, 864

\bibitem[\protect\citeauthoryear{{Gerssen}, {Kuijken} \&
  {Merrifield}}{{Gerssen} et~al.}{1999}]{gerssen99}
{Gerssen} J.,  {Kuijken} K.,    {Merrifield} M.~R.,  1999, MNRAS, 306, 926

\bibitem[\protect\citeauthoryear{{Gerssen}, {Kuijken} \&
  {Merrifield}}{{Gerssen} et~al.}{2003}]{gerssen2003}
{Gerssen} J.,  {Kuijken} K.,    {Merrifield} M.~R.,  2003, MNRAS, 345, 261

\bibitem[\protect\citeauthoryear{{Heller}, {Shlosman} \& {Englmaier}}{{Heller}
  et~al.}{2001}]{heller2001}
{Heller} C.~H.,  {Shlosman} I.,    {Englmaier} P.,  2001, ApJ, 553, 661

\bibitem[\protect\citeauthoryear{{Hern{\'a}ndez-Toledo},
  {Zendejas-Dom{\'{\i}}nguez} \& {Avila-Reese}}{{Hern{\'a}ndez-Toledo}
  et~al.}{2007}]{hernandeztoledo2007}
{Hern{\'a}ndez-Toledo} H.~M.,  {Zendejas-Dom{\'{\i}}nguez} J.,    {Avila-Reese}
  V.,  2007, AJ, 134, 2286

\bibitem[\protect\citeauthoryear{{Higdon}, {Buta} \& {Purcell}}{{Higdon}
  et~al.}{1998}]{higdon98}
{Higdon} J.~L.,  {Buta} R.~J.,    {Purcell} G.~B.,  1998, AJ, 115, 80

\bibitem[\protect\citeauthoryear{{Hunter}, {England}, {Gottesman}, {Ball} \&
  {Huntley}}{{Hunter} et~al.}{1988}]{hunter88}
{Hunter} J.~H.,  {England} M.~N.,  {Gottesman} S.~T.,  {Ball} R.,    {Huntley}
  J.~M.,  1988, ApJ, 324, 721

\bibitem[\protect\citeauthoryear{{Jarrett}, {Chester}, {Cutri}, {Schneider} \&
  {Huchra}}{{Jarrett} et~al.}{2003}]{jarrett2003}
{Jarrett} T.~H.,  {Chester} T.,  {Cutri} R.,  {Schneider} S.~E.,    {Huchra}
  J.~P.,  2003, AJ, 125, 525

\bibitem[\protect\citeauthoryear{{Jogee}, {Barazza}, {Rix}, {Shlosman},
  {Barden}, {Wolf}, {Davies}, {Heyer}, {Beckwith}, {Bell}, {Borch}, {Caldwell},
  {Conselice}, {Dahlen}, {H{\"a}ussler}, {Heymans}, {Jahnke}, {Knapen} \&
  {Laine}}{{Jogee} et~al.}{2004}]{jogee2004}
{Jogee} S.,  {Barazza} F.~D.,  {Rix} H.-W.,  {Shlosman} I.,  {Barden} M.,
  {Wolf} C.,  {Davies} J.,  {Heyer} I.,  {Beckwith} S.~V.~W.,  {Bell} E.~F.,
  {Borch} A.,  {Caldwell} J.~A.~R.,  {Conselice} C.~J.,  {Dahlen} T.,
  {H{\"a}ussler} B.,  {Heymans} C.,  {Jahnke} K.,  {Knapen} J.~H.,    {Laine}
  S.,  2004, ApJ, 615, L105

\bibitem[\protect\citeauthoryear{{Kenney}, {Carlstrom} \& {Young}}{{Kenney}
  et~al.}{1993}]{kenney93}
{Kenney} J. D.~P.,  {Carlstrom} J.~E.,    {Young} J.~S.,  1993, ApJ, 418, 687

\bibitem[\protect\citeauthoryear{{Kenney} \& {Lord}}{{Kenney} \&
  {Lord}}{1991}]{kenney91}
{Kenney} J. D.~P.,  {Lord} S.~D.,  1991, ApJ, 381, 118

\bibitem[\protect\citeauthoryear{{Kent}}{{Kent}}{1987}]{kent87}
{Kent} S.~M.,  1987, AJ, 93, 1062

\bibitem[\protect\citeauthoryear{{Kent} \& {Glaudell}}{{Kent} \&
  {Glaudell}}{1989}]{kent89}
{Kent} S.~M.,  {Glaudell} G.,  1989, AJ, 98, 1588

\bibitem[\protect\citeauthoryear{{Kranz}, {Slyz} \& {Rix}}{{Kranz}
  et~al.}{2003}]{kranz2003}
{Kranz} T.,  {Slyz} A.,    {Rix} H.,  2003, ApJ, 586, 143

\bibitem[\protect\citeauthoryear{{Laine}, {Shlosman}, {Knapen} \&
  {Peletier}}{{Laine} et~al.}{2002}]{laine2002}
{Laine} S.,  {Shlosman} I.,  {Knapen} J.~H.,    {Peletier} R.~F.,  2002, ApJ,
  567, 97

\bibitem[\protect\citeauthoryear{{Laurikainen} \& {Salo}}{{Laurikainen} \&
  {Salo}}{2002}]{laurikainen2002b}
{Laurikainen} E.,  {Salo} H.,  2002, MNRAS, 337, 1118

\bibitem[\protect\citeauthoryear{{Laurikainen}, {Salo} \& {Buta}}{{Laurikainen}
  et~al.}{2005}]{laurikainen2005}
{Laurikainen} E.,  {Salo} H.,    {Buta} R.,  2005, MNRAS, 362, 1319

\bibitem[\protect\citeauthoryear{{Laurikainen}, {Salo}, {Buta} \&
  {Knapen}}{{Laurikainen} et~al.}{2007}]{laurikainen2007}
{Laurikainen} E.,  {Salo} H.,  {Buta} R.,    {Knapen} J.~H.,  2007, MNRAS, 381,
  401

\bibitem[\protect\citeauthoryear{{Laurikainen}, {Salo}, {Buta} \&
  {Vasylyev}}{{Laurikainen} et~al.}{2004}]{laurikainen2004}
{Laurikainen} E.,  {Salo} H.,  {Buta} R.,    {Vasylyev} S.,  2004, MNRAS, 355,
  1251

\bibitem[\protect\citeauthoryear{{Lindblad} \& {Kristen}}{{Lindblad} \&
  {Kristen}}{1996}]{lindblad96b}
{Lindblad} P. A.~B.,  {Kristen} H.,  1996, A\&A, 313, 733

\bibitem[\protect\citeauthoryear{{Lindblad}, {Lindblad} \&
  {Athanassoula}}{{Lindblad} et~al.}{1996}]{lindblad96a}
{Lindblad} P. A.~B.,  {Lindblad} P.~O.,    {Athanassoula} E.,  1996, A\&A, 313,
  65

\bibitem[\protect\citeauthoryear{{Lynden-Bell}}{{Lynden-Bell}}{1979}]{lynden79}
{Lynden-Bell} D.,  1979, MNRAS, 187, 101

\bibitem[\protect\citeauthoryear{{Maciejewski}}{{Maciejewski}}{2006}]{maciejew%
ski2006}
{Maciejewski} W.,  2006, MNRAS, 371, 451

\bibitem[\protect\citeauthoryear{{Maciejewski} \& {Sparke}}{{Maciejewski} \&
  {Sparke}}{2000}]{maciejewski2000}
{Maciejewski} W.,  {Sparke} L.~S.,  2000, MNRAS, 313, 745

\bibitem[\protect\citeauthoryear{{Marinova} \& {Jogee}}{{Marinova} \&
  {Jogee}}{2007}]{marinova2007}
{Marinova} I.,  {Jogee} S.,  2007, ApJ, 659, 1176

\bibitem[\protect\citeauthoryear{{Martin}}{{Martin}}{1995}]{martin95b}
{Martin} P.,  1995, AJ, 109, 2428

\bibitem[\protect\citeauthoryear{{Martinez-Valpuesta}, {Knapen} \&
  {Buta}}{{Martinez-Valpuesta} et~al.}{2007}]{martinez2007}
{Martinez-Valpuesta} I.,  {Knapen} J.~H.,    {Buta} R.,  2007, AJ, 134, 1863

\bibitem[\protect\citeauthoryear{{Masset} \& {Tagger}}{{Masset} \&
  {Tagger}}{1997}]{masset97}
{Masset} F.,  {Tagger} M.,  1997, A\&A, 322, 442

\bibitem[\protect\citeauthoryear{{Men{\'e}ndez-Delmestre}, {Sheth},
  {Schinnerer}, {Jarrett} \& {Scoville}}{{Men{\'e}ndez-Delmestre}
  et~al.}{2007}]{menendez2007}
{Men{\'e}ndez-Delmestre} K.,  {Sheth} K.,  {Schinnerer} E.,  {Jarrett} T.~H.,
   {Scoville} N.~Z.,  2007, ApJ, 657, 790

\bibitem[\protect\citeauthoryear{{Merrifield} \& {Kuijken}}{{Merrifield} \&
  {Kuijken}}{1995}]{merrifield95}
{Merrifield} M.~R.,  {Kuijken} K.,  1995, MNRAS, 274, 933

\bibitem[\protect\citeauthoryear{{Michel-Dansac} \& {Wozniak}}{{Michel-Dansac}
  \& {Wozniak}}{2006}]{michel2006}
{Michel-Dansac} L.,  {Wozniak} H.,  2006, A\&A, 452, 97

\bibitem[\protect\citeauthoryear{{Miwa} \& {Noguchi}}{{Miwa} \&
  {Noguchi}}{1998}]{miwa98}
{Miwa} T.,  {Noguchi} M.,  1998, ApJ, 499, 149

\bibitem[\protect\citeauthoryear{{Mundell}, {Pedlar}, {Shone} \&
  {Robinson}}{{Mundell} et~al.}{1999}]{mundell99b}
{Mundell} C.~G.,  {Pedlar} A.,  {Shone} D.~L.,    {Robinson} A.,  1999, MNRAS,
  304, 481

\bibitem[\protect\citeauthoryear{{Navarro}, {Frenk} \& {White}}{{Navarro}
  et~al.}{1996}]{navarro96}
{Navarro} J.~F.,  {Frenk} C.~S.,    {White} S.~D.~M.,  1996, ApJ, 462, 563

\bibitem[\protect\citeauthoryear{{Patsis}}{{Patsis}}{2005}]{patsis2005}
{Patsis} P.~A.,  2005, MNRAS, 358, 305

\bibitem[\protect\citeauthoryear{{Patsis}, {Contopoulos} \&
  {Grosb{\o}l}}{{Patsis} et~al.}{1991}]{patsis91}
{Patsis} P.~A.,  {Contopoulos} G.,    {Grosb{\o}l} P.,  1991, A\&A, 243, 373

\bibitem[\protect\citeauthoryear{{Patsis}, {Skokos} \& {Athanassoula}}{{Patsis}
  et~al.}{2003}]{patsis2003}
{Patsis} P.~A.,  {Skokos} C.,    {Athanassoula} E.,  2003, MNRAS, 346, 1031

\bibitem[\protect\citeauthoryear{{Paturel}, {Petit}, {Prugniel}, {Theureau},
  {Rousseau}, {Brouty}, {Dubois} \& {Cambr{\'e}sy}}{{Paturel}
  et~al.}{2003}]{paturel2003}
{Paturel} G.,  {Petit} C.,  {Prugniel} P.,  {Theureau} G.,  {Rousseau} J.,
  {Brouty} M.,  {Dubois} P.,    {Cambr{\'e}sy} L.,  2003, A\&A, 412, 45

\bibitem[\protect\citeauthoryear{{Persic}, {Salucci} \& {Stel}}{{Persic}
  et~al.}{1996}]{persic96}
{Persic} M.,  {Salucci} P.,    {Stel} F.,  1996, MNRAS, 281, 27

\bibitem[\protect\citeauthoryear{{Puerari} \& {Dottori}}{{Puerari} \&
  {Dottori}}{1997}]{puerari97}
{Puerari} I.,  {Dottori} H.,  1997, ApJL, 476, L73

\bibitem[\protect\citeauthoryear{{Quillen}, {Frogel} \& {Gonzalez}}{{Quillen}
  et~al.}{1994}]{quillen94}
{Quillen} A.~C.,  {Frogel} J.~A.,    {Gonzalez} R.~A.,  1994, ApJ, 437, 162

\bibitem[\protect\citeauthoryear{{Rand} \& {Wallin}}{{Rand} \&
  {Wallin}}{2004}]{rand2004}
{Rand} R.~J.,  {Wallin} J.~F.,  2004, ApJ, 614, 142

\bibitem[\protect\citeauthoryear{{Rautiainen} \& {Salo}}{{Rautiainen} \&
  {Salo}}{1999}]{rautiainen99}
{Rautiainen} P.,  {Salo} H.,  1999, A\&A, 348, 737

\bibitem[\protect\citeauthoryear{{Rautiainen} \& {Salo}}{{Rautiainen} \&
  {Salo}}{2000}]{rautiainen2000}
{Rautiainen} P.,  {Salo} H.,  2000, A\&A, 362, 465

\bibitem[\protect\citeauthoryear{{Rautiainen}, {Salo} \& {Buta}}{{Rautiainen}
  et~al.}{2004}]{rautiainen2004}
{Rautiainen} P.,  {Salo} H.,    {Buta} R.,  2004, MNRAS, 349, 933

\bibitem[\protect\citeauthoryear{{Rautiainen}, {Salo} \&
  {Laurikainen}}{{Rautiainen} et~al.}{2002}]{rautiainen2002}
{Rautiainen} P.,  {Salo} H.,    {Laurikainen} E.,  2002, MNRAS, 337, 1233

\bibitem[\protect\citeauthoryear{{Rautiainen}, {Salo} \&
  {Laurikainen}}{{Rautiainen} et~al.}{2005}]{rautiainen2005}
{Rautiainen} P.,  {Salo} H.,    {Laurikainen} E.,  2005, ApJ, 631, L129

\bibitem[\protect\citeauthoryear{{Reese}, {Williams}, {Sellwood}, {Barnes} \&
  {Powell}}{{Reese} et~al.}{2007}]{reese2007}
{Reese} A.~S.,  {Williams} T.~B.,  {Sellwood} J.~A.,  {Barnes} E.~I.,
  {Powell} B.~A.,  2007, AJ, 133, 2846

\bibitem[\protect\citeauthoryear{{Regan} \& {Teuben}}{{Regan} \&
  {Teuben}}{2003}]{regan2003}
{Regan} M.~W.,  {Teuben} P.,  2003, ApJ, 582, 723

\bibitem[\protect\citeauthoryear{{Salo}}{{Salo}}{1991}]{salo91a}
{Salo} H.,  1991, A\&A, 243, 118

\bibitem[\protect\citeauthoryear{{Salo}, {Laurikainen}, {Rautiainen} \&
  {Buta}}{{Salo} et~al.}{2007}]{salo2007}
{Salo} H.,  {Laurikainen} E.,  {Rautiainen} P.,    {Buta} R.,  2007, in
  {Combes} F.,  {Palous} J.,  eds, IAU Symposium Vol.~235 of IAU Symposium,
  {Sticky-Particle Simulations of Barred Galaxies}.
pp 133--133

\bibitem[\protect\citeauthoryear{{Salo}, {Rautiainen}, {Buta}, {Purcell},
  {Cobb}, {Crocker} \& {Laurikainen}}{{Salo} et~al.}{1999}]{salo99}
{Salo} H.,  {Rautiainen} P.,  {Buta} R.,  {Purcell} G.~B.,  {Cobb} M.~L.,
  {Crocker} D.~A.,    {Laurikainen} E.,  1999, AJ, 117, 792

\bibitem[\protect\citeauthoryear{{Schwarz}}{{Schwarz}}{1981}]{schwarz81}
{Schwarz} M.~P.,  1981, ApJ, 247, 77

\bibitem[\protect\citeauthoryear{{Schwarz}}{{Schwarz}}{1984}]{schwarz84b}
{Schwarz} M.~P.,  1984, MNRAS, 209, 93

\bibitem[\protect\citeauthoryear{{Sellwood}}{{Sellwood}}{1981}]{sellwood81}
{Sellwood} J.~A.,  1981, A\&A, 99, 362

\bibitem[\protect\citeauthoryear{{Sellwood} \& {Sparke}}{{Sellwood} \&
  {Sparke}}{1988}]{sellwood88}
{Sellwood} J.~A.,  {Sparke} L.~S.,  1988, MNRAS, 231, 25P

\bibitem[\protect\citeauthoryear{{Sempere}, {Garcia-Burillo}, {Combes} \&
  {Knapen}}{{Sempere} et~al.}{1995}]{sempere95a}
{Sempere} M.~J.,  {Garcia-Burillo} S.,  {Combes} F.,    {Knapen} J.~H.,  1995,
  A\&A, 296, 45

\bibitem[\protect\citeauthoryear{{Sempere} \& {Rozas}}{{Sempere} \&
  {Rozas}}{1997}]{sempere97b}
{Sempere} M.~J.,  {Rozas} M.,  1997, A\&A, 317, 405

\bibitem[\protect\citeauthoryear{{Sheth}, {Elmegreen}, {Elmegreen}, {Capak},
  {Abraham}, {Athanassoula}, {Ellis}, {Mobasher}, {Salvato}, {Schinnerer},
  {Scoville}, {Spalsbury}, {Strubbe}, {Carollo}, {Rich} \& {West}}{{Sheth}
  et~al.}{2008}]{sheth2008}
{Sheth} K.,  {Elmegreen} D.~M.,  {Elmegreen} B.~G.,  {Capak} P.,  {Abraham}
  R.~G.,  {Athanassoula} E.,  {Ellis} R.~S.,  {Mobasher} B.,  {Salvato} M.,
  {Schinnerer} E.,  {Scoville} N.~Z.,  {Spalsbury} L.,  {Strubbe} L.,
  {Carollo} M.,  {Rich} M.,    {West} A.~A.,  2008, ApJ, 675, 1141

\bibitem[\protect\citeauthoryear{{Thompson}}{{Thompson}}{1981}]{thompson81}
{Thompson} L.~A.,  1981, ApJ, 244, L43

\bibitem[\protect\citeauthoryear{{Tremaine} \& {Weinberg}}{{Tremaine} \&
  {Weinberg}}{1984}]{tremaine84}
{Tremaine} S.,  {Weinberg} M.~D.,  1984, ApJ, 282, L5

\bibitem[\protect\citeauthoryear{{Treuthardt}, {Buta}, {Salo} \&
  {Laurikainen}}{{Treuthardt} et~al.}{2007}]{treuthardt2007}
{Treuthardt} P.,  {Buta} R.,  {Salo} H.,    {Laurikainen} E.,  2007, AJ, 134,
  1195

\bibitem[\protect\citeauthoryear{{Treuthardt}, {Salo}, {Rautiainen} \&
  {Buta}}{{Treuthardt} et~al.}{2008}]{treuthardt2008}
{Treuthardt} P.,  {Salo} H.,  {Rautiainen} P.,    {Buta} R.,  2008, submitted
  to AJ

\bibitem[\protect\citeauthoryear{{Tully}}{{Tully}}{1988}]{tully88}
{Tully} R.~B.,  1988, {Nearby galaxies catalog}.
Cambridge and New York, Cambridge University Press, 1988, 221 p.

\bibitem[\protect\citeauthoryear{{Vega Beltran}, {Zeilinger}, {Amico},
  {Schultheis}, {Corsini}, {Funes}, {Beckman} \& {Bertola}}{{Vega Beltran}
  et~al.}{1998}]{vega98}
{Vega Beltran} J.~C.,  {Zeilinger} W.~W.,  {Amico} P.,  {Schultheis} M.,
  {Corsini} E.~M.,  {Funes} J.~G.,  {Beckman} J.,    {Bertola} F.,  1998,
  A\&AS, 131, 105

\bibitem[\protect\citeauthoryear{{Weiner}, {Sellwood} \& {Williams}}{{Weiner}
  et~al.}{2001}]{weiner2001b}
{Weiner} B.~J.,  {Sellwood} J.~A.,    {Williams} T.~B.,  2001, ApJ, 546, 931

\bibitem[\protect\citeauthoryear{{Whyte}, {Abraham}, {Merrifield}, {Eskridge},
  {Frogel} \& {Pogge}}{{Whyte} et~al.}{2002}]{whyte2002}
{Whyte} L.~F.,  {Abraham} R.~G.,  {Merrifield} M.~R.,  {Eskridge} P.~B.,
  {Frogel} J.~A.,    {Pogge} R.~W.,  2002, MNRAS, 336, 1281

\bibitem[\protect\citeauthoryear{{Wilke}, {M{\"o}llenhoff} \&
  {Matthias}}{{Wilke} et~al.}{1999}]{wilke99}
{Wilke} K.,  {M{\"o}llenhoff} C.,    {Matthias} M.,  1999, A\&A, 344, 787

\bibitem[\protect\citeauthoryear{{Wilke}, {M{\"o}llenhoff} \&
  {Matthias}}{{Wilke} et~al.}{2000}]{wilke2000}
{Wilke} K.,  {M{\"o}llenhoff} C.,    {Matthias} M.,  2000, A\&A, 361, 507

\bibitem[\protect\citeauthoryear{{Zhang} \& {Buta}}{{Zhang} \&
  {Buta}}{2007}]{zhang2007}
{Zhang} X.,  {Buta} R.~J.,  2007, AJ, 133, 2584

\bibitem[\protect\citeauthoryear{{Zimmer}, {Rand} \& {McGraw}}{{Zimmer}
  et~al.}{2004}]{zimmer2004}
{Zimmer} P.,  {Rand} R.~J.,    {McGraw} J.~T.,  2004, ApJ, 607, 285

\end{thebibliography}

\end{document}